\documentclass[letterpaper,titlepage,11pt]{article}
\pdfoutput=1
\usepackage[colorlinks=true,urlcolor=blue,citecolor=magenta]{hyperref}
\usepackage{amssymb,amsmath,amsfonts}
\usepackage{epsfig}
\usepackage[sort&compress, merge]{natbib}
\usepackage{epsfig}
\usepackage{graphicx}
\usepackage{epstopdf}
\usepackage{caption}
\usepackage{subcaption}
\usepackage{amscd}
\usepackage{amsthm}
\usepackage{latexsym}
\usepackage{amsbsy}
\usepackage{bbm}
\usepackage[english]{babel}

\usepackage{tabularx}

\newcommand{\be}{\begin{eqnarray}}
\newcommand{\ee}{\end{eqnarray}}
\newcommand{\beq}{\begin{eqnarray}}
\newcommand{\eeq}{\end{eqnarray}}

\newcommand{\bk}{\textbf{k}}
\newcommand{\ds}{\textbf{ds}}
\newcommand{\dst}{\tilde{\textbf{ds}}}

\bibpunct{[}{]}{;}{n}{,}{,}

\def\clock{{\count0=\time
           \divide\count0 60
           \ifnum\count0<10 0\fi\the\count0
           \multiply\count0 -60 \advance\count0 \time
           :\ifnum\count0<10 0\fi \the\count0
         }}
\newcommand{\timestamp}{{\small\vbox{\hbox{\tt\jobname.tex}
\hbox{\the\day/\the\month/\the\year, \clock}}}}

\numberwithin{equation}{section}
\begin{document}

\begin{titlepage}
\ \ \vskip 1.8cm

\centerline{\Huge \bf New Geometries } 
\vskip 0.6cm
\centerline{\Huge \bf for Black Hole Horizons}

\vskip 1.6cm \centerline{\bf Jay Armas$^{1}$ and Matthias Blau$^{2}$} \vskip 0.7cm

\begin{center}
\sl $^{1}$ Physique Th\'{e}orique et Math\'{e}matique \\
Universit\'{e} Libre de Bruxelles and International Solvay Institutes \\
ULB-Campus Plaine CP231, B-1050 Brussels, Belgium\\
\end{center}
\vskip 0.3cm

\begin{center}
\sl $^{2}$ Albert Einstein Center for Fundamental Physics, University of Bern,\\
Sidlerstrasse 5, 3012 Bern, Switzerland
\end{center}
\vskip 0.3cm

\centerline{\small\tt jarmas@ulb.ac.be, blau@itp.unibe.ch}

\vskip 1.3cm \centerline{\bf Abstract} \vskip 0.2cm \noindent
We construct several classes of worldvolume effective actions for black holes by integrating out spatial sections of the worldvolume geometry of asymptotically flat black branes. This provides a generalisation of the blackfold approach for higher-dimensional black holes and yields a map between different effective theories, which we exploit by obtaining new hydrodynamic and elastic transport coefficients via simple integrations. Using Euclidean minimal surfaces in order to decouple the fluid dynamics on different sections of the worldvolume, we obtain local effective theories for ultraspinning Myers-Perry branes and helicoidal black branes, described in terms of a stress-energy tensor, particle currents and non-trivial boost vectors. We then study in detail and present novel compact and non-compact geometries for black hole horizons in higher-dimensional asymptotically flat space-time. These include doubly-spinning black rings, black helicoids and helicoidal $p$-branes as well as helicoidal black rings and helicoidal black tori in $D\ge6$.

\end{titlepage}

\tableofcontents

 %%%%%%%%%%%%%%%%%%%%%%%%%%%%%%%%%%%%%%
  %%%%%%%%%%%%%%%%%%%%%%%%%%%%%%%%%%%%%%
\section{Introduction}  \label{sec:intro}
Effective theories for black holes have been proven to be extremely useful in scanning and probing the properties of the highly intricate non-linear dynamics of higher-dimensional General Relativity. Amongst these are the blackfold approach \cite{Emparan:2009cs, Emparan:2009at} and the large D expansion \cite{Emparan:2013moa}, which have been used to find new black hole solutions as well as to study some of their properties, including stability. This, together with numerical methods \cite{Kleihaus:2012xh,Dias:2014cia,Kleihaus:2014pha, Figueras:2014dta} is slowly giving us a picture of the phase space of higher-dimensional black holes.

In particular, the blackfold approach has provided evidence for the existence of highly non-trivial vacuum black hole horizon geometries and topologies in higher-dimensional asymptotically flat space-time \cite{Emparan:2009vd, Armas:2015kra}. These include higher-dimensional (helical) black rings, helical black strings, black cylinders, black odd-spheres, products of black odd-spheres and higher-dimensional black helicoids. Some of these have also been generalised in the context of gravity with a cosmological constant \cite{Caldarelli:2008pz, Armas:2010hz}, in plane wave backgrounds \cite{Armas:2015kra} and with the addition of charges \cite{Caldarelli:2010xz, Emparan:2011hg}.

The blackfold approach is a long-wavelength effective theory for the dynamics of black branes, defined as a derivative expansion in the fields that characterise the brane. As originally developed in \cite{Emparan:2009cs, Emparan:2009at}, it consists of wrapping neutral black branes along arbitrary submanifolds in a background space-time. The dynamics of the brane are integrated out along the transverse directions to the brane worldvolume, leading to an effective theory characterised, to leading order, by a stress-energy tensor of the perfect fluid form. In this paper, following the study of minimal surfaces in \cite{Armas:2015kra}, we generalise this approach by further integrating out spatial subsections of worldvolume which have the geometry of Euclidean minimal surfaces. This allows us to construct new classes of asymptotically flat black holes such as higher-dimensional doubly-spinning black rings, helicoidal black rings and helicoidal black tori. 

\begin{figure}[h!] 
\centering
  \includegraphics[width=0.9\linewidth]{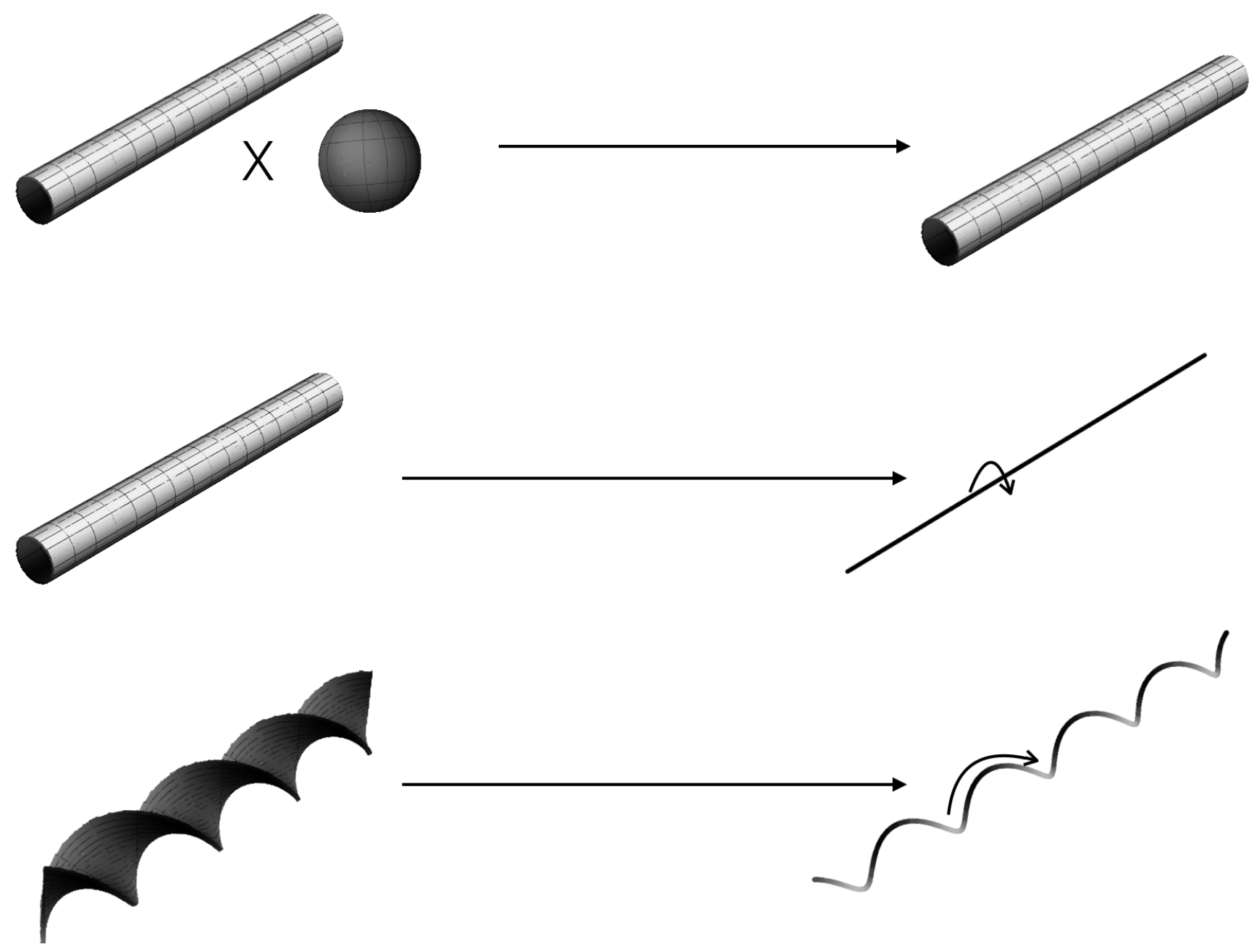}
  \begin{picture}(0,0)(0,0)
  \put(-400,270){ $(a) $}
\put(-300,270){ $ \mathbb{S}^{(n+1)} $}
\put(-200,270){ $ \int {\mathbb{S}^{(n+1)}} $}
  \put(-400,165){ $(b) $}
\put(-200,165){ $ \int {\mathbb{D}} $}
\put(-100,160){ $ \tilde{\mathcal{J}}^{a} $}
  \put(-400,65){ $(c) $}
\put(-200,65){ $ \int {\mathbb{I}} $}
\put(-100,65){ $ \tilde{v}^{a} $}
\end{picture} 
\caption{Schematic representation of integrating out degrees of freedom. In $(a)$ we depicted the usual blackfold approach consisting of integrating out the transverse $ \mathbb{S}^{(n+1)} $. In $(b)$ we integrate out the disc $ \mathbb{D}$ from a black cylinder geometry and obtain a Myers-Perry string with transverse angular momentum current. In $(c)$ we integrate out the width $\mathbb{I}$ of the helicoid and obtain a helical string with a boost vector carrying a linear momentum density.} \label{fig:eff}
\end{figure}
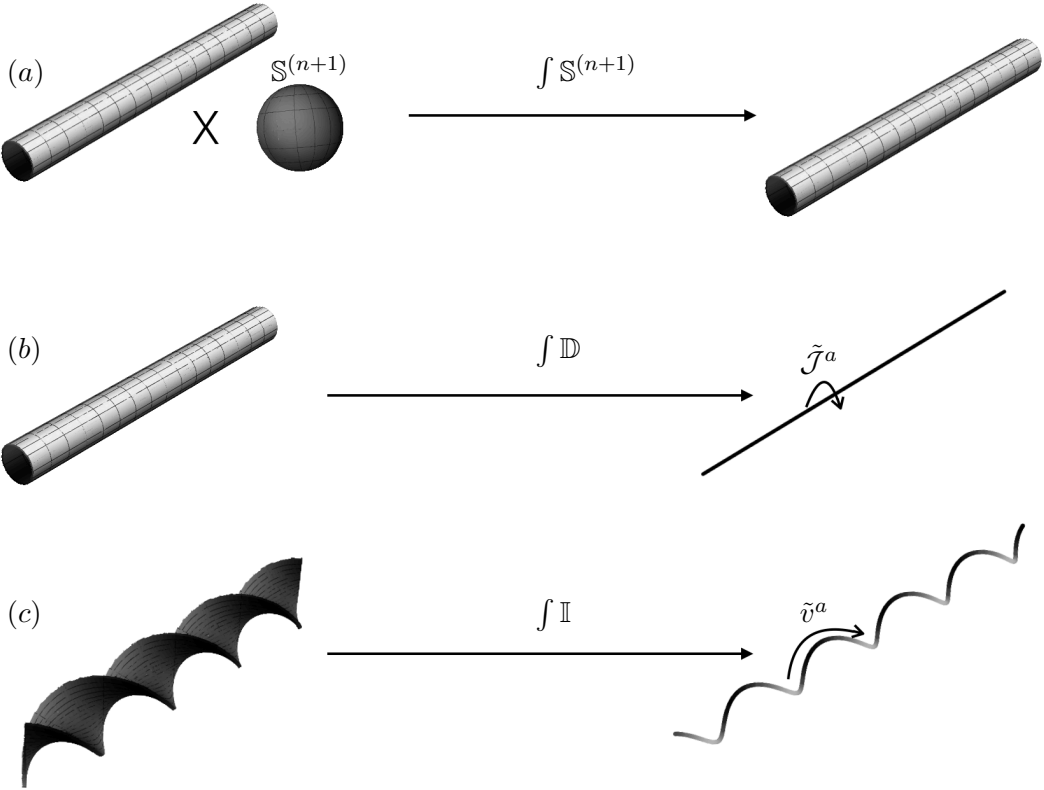

This generalised approach, which we will describe in detail in Sec.~\ref{sec:general}, provides a map between three different effective theories for black branes as portrayed in Fig.~\ref{fig:eff}. The original effective theory of \cite{Emparan:2009cs, Emparan:2009at} is obtained by integrating out the transverse $(n+1)$ angular coordinates, that is, integrating out the subspace $\mathbb{S}^{(n+1)}$. 
If one then considers the geometry of a string with a transverse disc of finite radius and integrates out the disc $\mathbb{D}$ one obtains the effective theory of a Myers-Perry string in the ultraspinning regime, characterised by a stress-energy tensor and a conserved particle current $\tilde{\mathcal{J}}^{a}$ carrying the transverse angular momentum in the plane of the disc. Instead, if one considers the geometry of the black helicoid found in \cite{Armas:2015kra} and integrates out the width of the helicoid $\mathbb{I}$ one obtains the effective theory of a helical black string, characterised by a stress-energy tensor and a boost vector $\tilde v^{a}$ carrying a linear momentum density. Both these theories will be obtained and studied in Secs.~\ref{sec:mpbranes}-\ref{sec:effhelicoids} and used to find new black hole geometries in Secs.~\ref{sec:blacktor}-\ref{sec:helicoidalrings}. In particular, we will show in Sec.~\ref{sec:hight} and Sec.~\ref{sec:hight2} how this map between effective theories can be used to obtain new transport coefficients for the different theories, given the transport coefficients of the original theory or known information about the resulting effective theories.

%%%%%%%%%%%%%%%%%%%%%%%%%%%%%%%%%%%%%%
%%%%%%%%%%%%%%%%%%%%%%%%%%%%%%%%%%%%%%
%%%%%%%%%%%%%%%%%%%%%%%%%%%%%%%%%%%%%%

\section{Generalized blackfold approach} \label{sec:general}
The blackfold approach, as introduced in \cite{Emparan:2009cs, Emparan:2009at}, consists of an effective theory for the dynamics of black branes obtained by integrating out the short-wavelength degrees of freedom of General Relativity. That is, if we associate the scales $\ell_a$ to short-wavelengths and $r_0$ to long-wavelengths, then the dynamics of the gravitational field $g_{\mu\nu}$ at distances $r\ll\ell_a$ is integrated out, giving rise to a long-wavelength effective theory valid for distances $r\gg r_0$, provided $r_0\ll\ell_a$.

For the case of neutral black branes, the effective geometry in the region $r_0\ll r\ll\ell_a$ where both short- and long-wavelength degrees of freedom interact, is locally given by the metric,
\beq \label{ds:blackp}
ds^2_p=\left(\gamma_{ab}(\sigma^{c})+\frac{r_0^n(\sigma^{c})}{r^n}u_{a}(\sigma^{c})u_{b}(\sigma_{c})\right)d\sigma^{a}d\sigma^{b}+\frac{dr^2}{1-\frac{r_0^n(\sigma^{c})}{r^n}}+r^2d\Omega_{(n+1)}^2+\dots~,
\eeq
where the \emph{dots} represent higher-order derivative corrections in the fields $\gamma_{ab},r_0,u^{a}$ with each derivative order being characterised by a suitable power of $\varepsilon=r_0/\ell_a$. The metric \eqref{ds:blackp} in $D=n+p+3$ dimensions is that of boosted neutral black brane of horizon radius $r_0$ and boost velocity $u^{a}$ which was allowed to vary along the coordinates $\sigma^{c}$ of the brane worldvolume $\mathcal{W}_{p+1}$. The worlvolume metric $\gamma_{ab}$ is the induced metric on $\mathcal{W}_{p+1}$, given by $\gamma_{ab}=g_{\mu\nu}\partial_a X^{\mu}\partial_b X^{\nu}$ where $X^{\mu}(\sigma^{c})$ is the set of mapping functions describing the position of the worldvolume in the background space-time.

At distances $r\gg r_0$, the gravitational field is sourced by the effective stress-energy tensor obtained from the brane metric \eqref{ds:blackp} using the Brown-York prescription \cite{Brown:1992br}. In the blackfold approach \cite{Emparan:2009cs, Emparan:2009at}, the dynamics in the angular directions in the $(n+1)$-sphere are, to leading order in $\varepsilon$, integrated out and hence play no role.\footnote{At higher orders, one can also integrate the transverse sphere and this will modify the effective theory such as to introduce elastic corrections in the stress-energy tensor \cite{Armas:2013hsa}.} The resultant stress-energy tensor is only defined along the $(p+1)$-worldvolume directions and takes the perfect fluid form
\beq \label{tbf}
T^{ab}=P\gamma^{ab}+(\epsilon+P)u^{a}u^{b}~~,~~u^{a}u_{a}=-1~~,
\eeq
where $P$ is the pressure and $\epsilon$ the energy density given by
\beq
P=-\frac{\Omega_{(n+1)}}{16\pi G}r_0^{n}~~,~~\epsilon=\frac{\Omega_{(n+1)}}{16\pi G}r_0^{n}(n+1)~~.
\eeq
These thermodynamic quantities satisfy the Gibbs-Duhem relation $\epsilon+P=\mathcal{T} s$, where $s$ is the local entropy density and $\mathcal{T}$ the local temperature and are given by
\beq \label{sbf}
s =\frac{\Omega_{(n+1)}}{4 G}r_0^{n+1}~~,~~\mathcal{T}=\frac{n}{4\pi r_0}~~.
\eeq

In order for gravitational objects of the form \eqref{ds:blackp} to be consistently coupled to gravity, and hence solve Einstein equations sourced by $T^{ab}$, the following leading order constraint equations must be satisfied \cite{Emparan:2007wm, Emparan:2009cs, Emparan:2009at, Camps:2010br,Camps:2012hw},
\beq \label{eq:bfeom}
\nabla_{a}T^{ab}=0~~,~~T^{ab}{K_{ab}}^{i}=0~~,
\eeq
where ${K_{ab}}^{i}={n^{i}}_{\mu}\nabla_{a}{u_{b}}^{\mu}$ is the extrinsic curvature tensor of the embedding geometry. Here we have defined the projector ${u_{b}}^{\mu}=\partial_{b}X^{\mu}$ along the $(p+1)$-worldvolume directions labeled by $a,b,c,...$ and the projector ${n^{i}}_{\mu}$ along the $(n+2)$-transverse directions to the worldvolume labeled by $i,j,k,...$ . In the presence of worldvolume boundaries, though this has not been shown directly from gravity but follows from probe brane considerations \cite{Emparan:2009cs, Emparan:2009at}, the blackfold equations \eqref{eq:bfeom} must be supplemented by the boundary condition
\beq \label{eq:bfeom1}
T^{ab}\eta_a|_{\mathcal{W}_{p+1}}=0~~,
\eeq
where $\eta^a$ is a unit normal vector to the brane boundary. Eq.~\eqref{eq:bfeom1} results in the requirement $r_0|_{\mathcal{W}_{p+1}}=0$.

\subsubsection*{The stationary sector}
If we focus on stationary solutions to the constraint equations \eqref{eq:bfeom}, then the hydrodynamic conservation equation $\nabla_{a}T^{ab}=0$ is solved if the fluid velocity $u^{a}$ is aligned with a worldvolume Killing vector field $\textbf{k}^{a}$, which we take to be of the following general form
\be
\bk^{a}\partial_a=\partial_\tau+\Omega^{a}\partial_{\phi_a}~~,
\eeq
where $\tau$ labels the time-like direction of the worldvolume and $\Omega^{a}$ denotes the set of angular velocities along each of the Cartan angles $\phi_a$ of the worldvolume geometry. Furthermore, the local temperature $\mathcal{T}$ must be redshifted such that $\mathcal{T}=T/\bk$ where $T$ is the global temperature and $\bk=|-\gamma_{ab}\bk^{a}\bk^{b}|^{1/2}$ is the modulus of the Killing vector field $\textbf{k}^{a}$. 

In this case, the blackfold equations \eqref{eq:bfeom} and the boundary condition \eqref{eq:bfeom1} reduce to
\beq \label{eq:bfeom2}
T^{ab}{K_{ab}}^{i}=0~~,~~\bk|_{\mathcal{W}_{p+1}}=0~~,
\eeq
and we can integrate them to an effective free energy \cite{Emparan:2009at},
\beq\label{eq:free1}
\mathcal{F}[X^{i}]=-\int_{\mathcal{B}_p}R_0 dV_{(p)}P~~,
\eeq
where $\mathcal{B}_p$ is the spatial part of the worldvolume and we have assumed that the worldvolume Killing vector is hypersurface orthogonal such that $d^{p+1}\sigma\sqrt{-\gamma}=R_0dV_{(p)}$ with $R_0$ being the modulus of $\partial_\tau$. The leading order free energy \eqref{eq:free1} is the integral over the worldvolume of the local Gibbs free energy density $\mathcal{G}=-P$. 
The free energy \eqref{eq:free1} obeys the relation
\beq\label{thermo1}
\mathcal{F}=M-T S-\sum_{a}\Omega^{a}J^{a}~~,
\eeq
where $M$ is the total mass, $S$ is the total entropy and $J^{a}$ is the angular momentum associated with each Cartan angle $\phi_a$. These can be obtained using the expressions \cite{Armas:2014rva},
\beq \label{thermo2}
S=-\frac{\partial \mathcal{F}}{\partial T}~~,~~J^{a}=-\frac{\partial \mathcal{F}}{\partial \Omega^{a}}~~.
\eeq
Configurations that satisfy \eqref{eq:bfeom2} also satisfy a Smarr-type relation of the form
\beq \label{eq:smarr}
(n+p)M-(n+p+1)(TS+\sum\Omega^{a}J^{a})=\boldsymbol{\mathcal{T}}~~,
\eeq
where $\boldsymbol{\mathcal{T}}$ is the total integrated tension given by
\beq \label{eq:tension}
\boldsymbol{\mathcal{T}}=-\int_{\mathcal{B}_p}dV_{(p)}R_0\left(\gamma^{ab}+\xi^{a}\xi^{b}\right)T_{ab}~~,
\eeq
and $\xi^{a}\partial_a=\partial_\tau$. If we are dealing with compact black hole configurations in flat space-time then we must have that $\boldsymbol{\mathcal{T}}=0$.

\subsection{Integrating out the worldvolume geometry} \label{sec:int}
As mentioned above, in the usual blackfold approach, one integrates out the angular directions in the transverse $(n+1)$-sphere since the dynamics of the brane, to leading order, along those directions and along the worldvolume decouple from each other. However, in certain cases, one can further integrate spatial sections of the worldvolume geometry obtaining other effective theories and effective free energies (or effective actions).\footnote{Implicitly, the idea of novel effective theories in the presence of other length scales can be found in \cite{BlancoPillado:2007iz, Emparan:2010sx}.} 

Consider the case in which the spatial part of the worldvolume is a product space such that $\mathcal{B}_p=\mathcal{B}_{p-(w+m)}\times \mathbb{E}_{M}^{(w+m)}$ with $w,m>1$. In this case, the dynamics of the effective fluid on $\mathcal{B}_p$ decouples and hence its motion is independent on both spaces $\mathcal{B}_{p-(w+m)}$ and $\mathbb{E}_{M}^{(w+m)}$ (or in an $m$-dimensional subspace $\tilde{\mathbb{E}}_{M}^{(m)}$ of $\mathbb{E}_{M}^{(w+m)}$) , if the extrinsic (elastic) equation
\beq \label{ext:1t}
T^{\hat a \hat b}{K_{\hat a \hat b}}^{i}=0~~,~~T^{a \hat b}{K_{a \hat b}}^{i}=0~~,
\eeq
along the $m$-directions of the space $\mathbb{E}_{M}^{(w+m)}$, which are labeled by the indices $\hat a, \hat b,\hat c,...$, is satisfied together with the validity requirement $r_0\ll \ell_{\hat a}$. Then one can integrate out the dynamics on the $m$-dimensional subspace $\tilde{\mathbb{E}}_{M}^{(m)}$, resulting in an effective theory for the worldvolume $\tilde{\mathcal{W}}_{\tilde p+ 1}$ and with spatial sections $\tilde{\mathcal{B}}_{\tilde p}$ where $\tilde p=p-m$. The worldvolume has $(n+m+2)$-transverse directions and lives in $D=\tilde n+\tilde p+3$-dimensional space-time, where $n=\tilde n-m$ and $\tilde n>m$. Performing this integration directly in the leading order free energy \eqref{eq:free1} leads to
\beq \label{eq:free1t}
\tilde{\mathcal{F}}[X^{i}]=-\int_{\tilde{\mathcal{B}}_{\tilde p}}\tilde {R}_0 d\tilde{V}_{(\tilde{p})}\tilde P+\mathcal{O}(\varepsilon)~~,
\eeq
where we have defined the effective pressure $\tilde P$ as
\beq
\tilde P=\int_{\tilde{\mathbb{E}}_{M}^{(m)}}d^{m}\sigma\sqrt{\gamma_{m}}P~~,
\eeq
with $\gamma_m$ being the determinant of the metric on $\tilde{\mathbb{E}}_{M}^{(m)}$. We denote all quantities of the resultant geometry with a \emph{tilde}, and hence we have a map $\{\gamma_{ab},u^{a},\bk\} \to \{\tilde \gamma_{ab},\tilde u^{a},\tilde \bk,...\} $, where the $\emph{dots}$ here may contain other fields that result from the integration.

Since the resultant effective free energy \eqref{eq:free1t} depends now on the induced metric $\tilde \gamma_{ab}=g_{\mu\nu}\partial_a \tilde X^{\mu}\partial_b \tilde X^{\nu}$, one may readily obtain the effective stress-energy tensor $\tilde T^{ab}$ using
\beq
\tilde T^{ab}=\frac{2}{\sqrt{-\tilde\gamma}}\frac{\delta \tilde{\mathcal{F}}}{\delta \tilde \gamma_{ab}}~~,
\eeq
as well as other thermodynamic quantities/currents using \eqref{thermo1}-\eqref{thermo2} and corresponding local versions. From here, and with a few educated guesses, one may deduce the leading order effective theory, which, when restricted to the stationary sector, leads to \eqref{eq:free1t}.

Large families of product spaces $\mathcal{B}_{p-(w+m)}\times \mathbb{E}_{M}^{(w+m)}$, for which no coupling occurs between the dynamics, exist, in particular when $\mathbb{E}_{M}^{(w+m)}$ is a compact Euclidean minimal surface that satisfies \eqref{ext:1t} along, at least, $m$-directions. These surfaces are characterised by having zero mean extrinsic curvature $K^{i}=\gamma^{ab}{K_{ab}}^{i}=0$ and if $(w+m)=1,2$, these have been classified in \cite{Armas:2015kra} and consist of discs and helicoids, while if $(w+m)>2$ it was shown that higher-dimensional versions of discs (even-balls) and helicoids also satisfy \eqref{ext:1t}. Based on these geometries we will give examples of two distinct worldvolume effective theories obtained by integrating Euclidean minimal surfaces $\mathbb{E}_{M}^{(m)}$ (in which case we set $w=0$) or subspaces $\tilde{\mathbb{E}}_{M}^{(m)}$ of $(w+m)$-dimensional Euclidean minimal surfaces:

\begin{itemize}

\item In Sec.~\ref{sec:mpbranes} we consider the effective theory that results from integrating out discs of finite size $\ell_{\hat a}$ and its higher-dimensional version (even-balls). In this case the submanifold $\mathbb{E}_{M}^{(m)}$ (where we have set $w=0$) is simply $m$-dimensional Euclidean space $\mathbb{R}^{(m)}$ with an ellipsoidal boundary where $m$ is even dimensional. This bounded Euclidean $\mathbb{R}^{(m)}$ space can be split into a product of two-planes with a circular boundary, each with an associated lengscale $\ell_{\hat{a}}=\tilde \omega^{-1}_{\hat a}$. Each of these discs is characterised by a 1-parameter family of $U(1)$ isometries with closed orbits and hence integrating them out results in an effective theory for flat branes with induced metric $\tilde\gamma_{ab}$, fluid velocity $\tilde u^{a}$ and a set of $m/2$ local angular velocities $\tilde\omega_{\hat a}$ associated to the corresponding set of transverse angular momenta $\tilde J_{\perp}^{a}$ in the sense defined in \cite{Armas:2013hsa, Armas:2014rva}.

\item In Sec.~\ref{sec:effhelicoids} we obtain the effective theory that results from integrating out finite size intervals with length $\ell_{\hat{a}}=\tilde \omega^{-1}_{a}$. In this case the submanifold $\mathbb{E}_{M}^{(w+m)}$ is an $(w+m)$-dimensional helicoid geometry characterised by a 1-parameter family of $U(1)$ isometries whose orbits are not closed. The underlying effective theory, obtained after integrating an $(w+m-1)$ section of $\mathbb{E}_{M}^{(w+m)}$, is that of a flat brane with induced metric $\tilde\gamma_{ab}$, fluid velocity $\tilde u^{a}$ and a non-trivial boost vector $\tilde v^{a}$ carrying a linear momentum density. If $(w+m)=2$ and hence $\mathbb{E}_{M}^{(2)}$ is a two-dimensional helicoid geometry, we integrate out a finite line segment (the width of the helicoid) and obtain a helical string. 

\end{itemize}

\subsection{The higher-order effective free energy functional}
In \eqref{eq:free1} we have given the leading order effective free energy. However, higher-derivative corrections can be accounted for. In general one has a series of the form
\beq
\mathcal{F}[X^{i}]=\mathcal{F}[X^{i}]_{(0)}+\mathcal{F}[X^{i}]_{(1)}+\mathcal{F}[X^{i}]_{(2)}+\dots~~,
\eeq
where the subscript $k$ in the corrections $\mathcal{F}[X^{i}]_{(k)}$ indicates the order $\varepsilon^{k}$ of the derivative expansion. The procedure of integrating out the geometry, explained in the previous section, is an order-by-order procedure. To order $\mathcal{O}(\varepsilon^2)$, the effective free energy functional for the neutral branes \eqref{ds:blackp} is given by \cite{Armas:2013hsa, Armas:2013goa}
\beq\label{eq:free}
\begin{split}
\mathcal{F}[X^{i}]=-\int_{\mathcal{B}_{p}} R_0dV_{(p)}\Big(P&+\upsilon_1\mathfrak{a}^{c}\mathfrak{a}_c+\upsilon_2 \mathcal{R}+\upsilon_{3}u^{a}u^{b}\mathcal{R}_{ab} \\
&+\lambda_1 K^{i}K_{i}+\lambda_2 K^{abi}K_{abi}+\lambda_3 u^{a}u^{b}{K_{a}}^{ci}K_{bci}\Big)~~,
\end{split}
\eeq
and is valid for $n>2$, as otherwise one would need to take into account backreaction effects. In \eqref{eq:free} we have introduced the fluid acceleration $\mathfrak{a}^{c}=u^{a}\nabla_au^{c}$, the worldvolume Ricci scalar $\mathcal{R}$ and the worldvolume Ricci tensor $\mathcal{R}_{ab}$.\footnote{The free energy \eqref{eq:free} obtained in \cite{Armas:2013hsa, Armas:2013goa}, as well as other studies available in the literature for space-filling fluids, does not deal with boundaries. In the presence of boundaries, which is the case of the configurations we study here, the invariant $\omega_{ab}\omega^{ab}$, where $\omega_{ab}$ is the fluid vorticity, may be independent of the scalars $\mathfrak{a}^{c}\mathfrak{a}_c$ and $u^{a}u^{b}\mathcal{R}_{ab}$ and hence must be added to \eqref{eq:free}. However, for all cases studied in this paper we have that $\omega_{ab}=0$.} The transport coefficients, $\lambda_1,\lambda_2,\lambda_3$ were obtained in \cite{Armas:2013hsa} via the local computations of \cite{Armas:2011uf, Camps:2012hw} and read
\beq
\lambda_1=-Pr_0^{2}\frac{3n+4}{2n^2(n+2)}\xi(n)~~,~~\lambda_2=-Pr_0^{2}\frac{1}{2(n+2)}\xi(n)~~,~~\lambda_3=-Pr_0^{2}\xi(n)~~,
\eeq
where we have defined the function $\xi(n)$,
\beq
\xi(n)=\frac{n\tan(\pi/n)}{\pi}\frac{\Gamma\left(\frac{n+1}{n}\right)^{4}}{\Gamma\left(\frac{n+2}{n}\right)^{2}}~~,~~n>2~~.
\eeq
The remaining transport coefficients $\upsilon_1,\upsilon_2,\upsilon_3$ have not yet been determined and require pushing the analysis of \cite{Emparan:2007wm, Armas:2011uf, Camps:2012hw} to next order. However, as we will see below, we have found a simpler way of obtaining $\upsilon_1$. 

Performing the integration over $\mathbb{E}_{M}^{(m)}$ leads to
\beq\label{eq:freet}
\begin{split}
\tilde{\mathcal{F}}[\tilde{X}^{i}]=-\int_{\tilde{\mathcal{B}}_{\tilde p}} \tilde R_0d\tilde V_{(\tilde p)}\Big(\tilde P&+\tilde\upsilon_1\tilde{\mathfrak{a}}^{c}\tilde{\mathfrak{a}}_c+\tilde{\upsilon}_2 \tilde{\mathcal{R}}+\tilde \upsilon_{3}\tilde u^{a}\tilde u^{b}\tilde{\mathcal{R}}_{ab} \\
&+\tilde\lambda_1 \tilde K^{i}\tilde K_{i}+\tilde \lambda_2 \tilde K^{abi}\tilde K_{abi}+\tilde \lambda_3 \tilde u^{a}\tilde u^{b}{\tilde K_{a}}^{ci}\tilde K_{bci}+\dots\Big)~~,
\end{split}
\eeq
where the \emph{dots} may contain terms which are interpreted as corrections $\tilde P_{(2)}$ to the leading order pressure $\tilde P$, or terms involving derivatives of boost vectors $\tilde v^{a}$, or even terms which involve couplings to currents of transverse angular momentum to first order in derivatives, even though the free energy \eqref{eq:free} originally only contained second order corrections. This shows that this procedure allows to extract the transport coefficients $\tilde \upsilon_i,\tilde \lambda_i$ via a simple integration, given $ \upsilon_i,\lambda_i$, by studying configurations with non-zero geometric invariants. As will be explained in Sec.~\ref{sec:hight}, this has allowed us to determine $\upsilon_1$ in the form\footnote{The result \eqref{up1} holds when the fluid vorticity $\omega_{ab}$ vanishes.}
\beq \label{up1}
\upsilon_1=-\frac{n}{2}Pr_0^{2}~~,
\eeq
with $n>2$, which we will use in order to asses higher-order corrections to doubly-spinning black rings in Sec.~\ref{sec:blacktor}.

\subsubsection*{Validity of the approach}
In \cite{Armas:2015kra} we have put forth an order-by-order method for analysing the regime of validity of the blackfold approach for any given configuration based on the effective free energy functional \eqref{eq:free}. Using the fact that all transport coefficients $\upsilon_i,\lambda_i$ scale as $r_0^{n+2}\propto \bk^{n+2}$ from \eqref{eq:free} we must have the requirements \cite{Armas:2015kra},
\beq\label{eq:req}
r_0\ll\left(|\mathfrak{a}^{c}\mathfrak{a}_c|^{-\frac{1}{2}},|\mathcal{R}|^{-\frac{1}{2}},|u^{a}u^{b}\mathcal{R}_{ab}|^{-\frac{1}{2}},|K^{i}K_{i}|^{-\frac{1}{2}},|K^{abi}K_{abi}|^{-\frac{1}{2}},|u^{a}u^{b}{K_{a}}^{ci}K_{bci}|^{-\frac{1}{2}}\right)~~.
\eeq
In the case of flat space-time, because of the of Gauss-Codazzi equations it is not necessary to analyse the last two invariants, as they can be expressed as a combination of the others \cite{Armas:2013hsa}. When performing the integration as in \eqref{eq:freet} exactly the same type of requirements hold, where now all quantities should have \emph{tildes}. We note that, as explained in \cite{Armas:2015kra}, for configurations with boundaries for which $\bk=0$ at some point in the worldvolume due to \eqref{eq:bfeom2}, one can verify that $|\mathfrak{a}^{c}\mathfrak{a}_c|^{-\frac{1}{2}}\propto\bk^{2}$ and hence it is not possible to satisfy the requirement $r_0\ll|\mathfrak{a}^{c}\mathfrak{a}_c|^{-\frac{1}{2}}$ close to the boundary. As noted in \cite{Armas:2015kra}, it is expected that the effective theory breaks down close to the boundary and one must assume the existence of a well defined boundary expansion that refines the approximation order-by-order. In this paper, we will assume that this is the case and only study the validity requirements away from the boundaries.

%%%%%%%%%%%%%%%%%%%%%%%%%%%%%%%%%%%%%%
%%%%%%%%%%%%%%%%%%%%%%%%%%%%%%%%%%%%%%
%%%%%%%%%%%%%%%%%%%%%%%%%%%%%%%%%%%%%%
%%%%%%%%%%%%%%%%%%%%%%%%%%%%%%%%%%%%%%
\section{Effective theory for ultraspinning Myers-Perry branes} \label{sec:mpbranes}
In this section we give an example of an effective theory obtained by integrating out the geometry of a disc or of even-balls, for which the resulting local black brane metric is that of ultraspinning Myers-Perry branes. This effective theory can also potentially be obtained as a limit of the blackfold effective theory for Myers-Perry branes with arbitrary angular momenta. In Sec.~\ref{sec:blacktor} we construct doubly-spinning black rings using this effective theory.

\subsection{Effective theory for ultraspinning Myers-Perry strings: an example} \label{sec:effMP}
Here we consider the simplest example of the type of construction that we have advertised Sec.~\ref{sec:int}. First we review the blackfold construction of a singly-spinning Myers-Perry black hole in the ultraspinning regime \cite{Emparan:2009vd}. This is described by the induced geometry,
\beq \label{ds:disc}
\textbf{ds}^2=-d\tau^2+d\rho^2+\rho^2d\phi^2~~,
\eeq
where $\ds^2=\gamma_{ab}d\sigma^{a}d\sigma^{b}$. The worldvolume geometry is rotating with angular velocity $\Omega$ such that its worldvolume Killing vector field and respective modulus are given by
\beq
\textbf{k}^{a}\partial_a=\partial_\tau+\Omega\partial_\phi~~,~~\textbf{k}^2=1-\Omega^2\rho^2~~.
\eeq 
This stationary geometry is a Lorentzian minimal surface satisfying $K^{i}=0$ since ${K_{ab}}^{i}=0$ and hence trivially solves the blackfold equations \eqref{eq:bfeom2}. Furthermore, according to the blackfold equations \eqref{eq:bfeom2}, the geometry acquires a boundary at $\rho=\rho_+=\Omega^{-1}$ where $\bk=0$ which must satisfy $r_+\ll\rho_+$ \cite{Armas:2015kra}. Therefore, spatial sections of the worldvolume have the geometry of a disc of radius $\rho_+$ while the topologies of the black hole horizons that they give rise to are spherical $\mathbb{S}^{(D-2)}$. This describes the geometry of Myers-Perry black holes in the ultraspinning regime \cite{Emparan:2009vd}.

Clearly, we can add an extra flat direction $z$ to the geometry \eqref{ds:disc} and boost it with boost velocity $\beta$ while still satisfying the blackfold equations \eqref{eq:bfeom2}. This is due to the fact that the resulting geometry is still minimal and satisfies ${K_{ab}}^{i}=0$. It is, therefore, a string of discs (a filled cylinder) with line element and Killing vector field
\beq \label{ds:sdisc}
\textbf{ds}^2=-d\tau^2+d\rho^2+\rho^2d\phi^2+dz^2~~,~~\textbf{k}^a\partial_a=\partial_\tau+\Omega\partial_\phi+\beta\partial_z~~,
\eeq
and describes a boosted string of ultraspinning Myers-Perry black holes with topology $\mathbb{R}\times\mathbb{S}^{(D-3)}$.  Due to the presence of the non-zero boost $\beta$, the size of the disc is finite and equal to $\rho_+=\sqrt{1-\beta^2}\Omega^{-1}$ while the size of the $z$-direction is, obviously, infinite. Hence, the spatial worldvolume geometry has two widely separated length scales for which the dynamics decouple on $\mathcal{B}_p=\mathbb{R}\times \mathbb{E}^{2}_M$, where $\mathbb{E}^{2}_M$ is a two-plane with a circular boundary, hence with the topology of a disc. Therefore, we can describe this geometry effectively by integrating out the disc section of \eqref{ds:sdisc} and obtaining the effective geometry of a boosted string
\beq \label{eq:effgeo}
\dst^2=-d\tau^2+dz^2~~,~~\tilde\bk^{a}\partial_a=\partial_\tau+\beta\partial_z
\eeq
together with a conserved particle current $\tilde{\mathcal{J}}^{a}=\tilde{\mathcal{J}}\tilde u^{a}$ carrying the transverse angular momentum, where $\tilde{\mathcal{J}}$ is a density of transverse angular momentum. The ultraspinning Myers-Perry string has two equivalent descriptions: either as a filled cylinder ($p=3$) described by a local boosted Schwarzschild brane metric or as a string ($p=1$) with a conserved particle current described locally by a boosted Myers-Perry brane in the ultraspinning regime. The reason for this may be understood by looking at the spinning action \cite{Armas:2014rva} for Myers-Perry $p$-branes with arbitrary angular momenta and then taking the ultraspinning limit. In this limit two directions of the transverse $(n+1)$-sphere blow up and the source can be effectively described by a $p+2$ brane. 

\subsubsection*{Obtaining the underlying local effective theory}
The procedure of integrating out the disc modifies the thermodynamic properties of the black brane. Since we are dealing with stationary configurations we can obtain the properties and effective action for this effective string by evaluating its free energy using \eqref{eq:free},
\beq\label{eq:freestring}
\begin{split}
\mathcal{F}&=\frac{\Omega_{(n+1)}}{16\pi G} r_+^{n}\int dz\int_{0}^{2\pi}d\phi\int_{0}^{\rho_+}d\rho~\rho \left(1-\Omega^2\rho^2-\beta^2\right)^{\frac{n}{2}} \\
&=\frac{\Omega_{(n+1)}}{16\pi G} r_+^{n}\frac{2\pi}{(n+2)\Omega^2}\int dz\left(1-\beta^2\right)^{\frac{n+2}{2}}~~.
\end{split}
\eeq 
Using the effective geometry \eqref{eq:effgeo} we can rewrite the free energy \eqref{eq:freestring} as 
\beq\label{eq:freestring1}
\tilde{\mathcal{F}}[\tilde X^{i}]=\frac{\Omega_{(\tilde n+1)}}{16\pi G}\frac{\tilde r_+^{\tilde n -2}}{\tilde\Omega^2}\int_{\mathcal{B}_{\tilde p}} \sqrt{-\tilde \gamma}~ \tilde{\bk}^{\tilde n}~~,
\eeq
where, since we have integrated a disc, we have used that $n=\tilde n-2$ and $D=\tilde n+\tilde p+3$ with $\tilde n>2$ and $\tilde p=1$ in this case. We note that we must have $D\ge7$, which is consistent with the fact that Myers-Perry black holes only exhibit ultraspinning regimes in $D\ge6$. Furthermore, we made the following redefinitions $\Omega\to\tilde\Omega$ and $r_+\to\tilde r_+$. The effective free energy \eqref{eq:freestring1} describes the dynamics of stationary configurations composed locally of \eqref{ds:sdisc} and can now be used to construct new geometries by wrapping the effective string on a one-dimensional spatial submanifold. 

However, one may wonder what is the underlying effective theory which, when restricted to the stationary sector, gives rise to an effective free energy of the form \eqref{eq:freestring1}. There is no established procedure in order to extract the underlying theory from the free energy \eqref{eq:freestring1} and it involves guesswork and clues that can be obtained from \eqref{eq:freestring1}, such as the stress-energy tensor. In particular, the free energy \eqref{eq:freestring} can be interpreted as the integral over the worldvolume of the local effective Gibbs free energy density of the string. Therefore, consider the following local Gibbs free energy
\beq \label{gibbsguess}
\tilde{\mathcal{G}}=\frac{\Omega_{(\tilde n+1)}}{16\pi G} \frac{\tilde r_0^{\tilde n-2}}{\tilde{\omega}^2}~~,~~\tilde r_0=\frac{(\tilde n-2)}{4\pi\tilde{\mathcal{T}}}~~,
\eeq
where $\tilde{\mathcal{T}}$ is the local temperature and $\tilde\omega$ the local density of transverse angular velocity. The guess \eqref{gibbsguess} for the form of the Gibbs free energy was motivated from the form of \eqref{eq:freestring1} in which the effective free energy functional for stationary configurations appears to be equivalent to \eqref{gibbsguess} when taking into account the local Lorentz factor $\tilde{\textbf{k}}$ on $\mathcal{W}_{\tilde p+1}$. When writing \eqref{gibbsguess}, we assume that $\tilde \omega\tilde r_0\ll1$ since we must preserve the original validity regime prior to the integration of the disc. Using a local version of the thermodynamic relations \eqref{thermo1}-\eqref{thermo2}, and furthermore that $\tilde P=-\tilde{\mathcal{G}}$, one finds the string effective pressure, entropy density and particle current density
\beq \label{p:mp}
\tilde P=-\frac{\Omega_{(\tilde n+1)}}{16\pi G} \frac{\tilde r_0^{\tilde n-2}}{\tilde{\omega}^2}~~,~~\tilde s=\frac{\Omega_{(\tilde n+1)}}{4 G} \frac{\tilde r_0^{\tilde n-1}}{\tilde{\omega}^2}~~,~~\tilde{\mathcal{J}}=\frac{\Omega_{(\tilde n+1)}}{8 \pi G} \frac{\tilde r_0^{\tilde n-2}}{\tilde{\omega}^3}~~.
\eeq
In fact, these quantities can be obtained by using the results of \cite{Armas:2011uf} for Myers-Perry branes with arbitrary angular momentum and then taking the ultraspinning limit. In order to obtain the energy density $\tilde\epsilon$ we vary \eqref{eq:freestring1} with respect to the metric $\tilde\gamma_{ab}$ and obtain the effective fluid stress-energy tensor, given by
\beq
\tilde T^{ab}=\tilde P\tilde{\gamma}^{ab}-\tilde n\tilde P\tilde {u}^{a}\tilde {u}^{b}~~,~~\tilde u^{a}=\tilde\bk^{a}/\tilde\bk~~.
\eeq
Since the stress-tensor takes the perfect fluid form, from here we conclude that $\tilde{\epsilon}+\tilde P=-n\tilde P$. It is then easy to verify that that these thermodynamic densities satisfy local Euler-Gibbs-Duhem relations for a fluid carrying particle charge 
\beq
\tilde{\epsilon}+\tilde P=\tilde{\mathcal{T}}\tilde s+\tilde{\omega}\tilde{\mathcal{J}}~~,~~d\tilde P=\tilde sd\tilde{\mathcal{T}}+\tilde{\mathcal{J}}d\tilde\omega~~,
\eeq 
as well as the first law $d\tilde \epsilon=\tilde{\mathcal{T}}d\tilde s+\tilde\omega d\tilde{\mathcal{J}}$. Here, the local transverse angular velocity $\tilde\omega$ plays the role of a chemical potential. Promoting all the thermodynamic potentials $\tilde{\mathcal{T}},\tilde{\omega}$ and the metric $\tilde\gamma_{ab}$ as well as the boost velocity $\tilde u^{a}$ to functions of the worldvolume coordinates $\sigma^{a}$ finalises the effective theory.

If we focus on stationary configurations, then we must solve the conservation equation \eqref{eq:bfeom}, assuming that both entropy and transverse angular momentum currents are conserved. This requires, as in \cite{Caldarelli:2010xz}, that $\tilde {u}^{a}$ is aligned with a Killing vector field of the worldvolume geometry $\tilde{\bk}^{a}$ such that $\tilde {u}^{a}=\tilde{\bk}^{a}/\tilde{\bk}$ and furthermore that global potentials describing the temperature $\tilde T$ and transverse angular velocity $\tilde \Omega$ are given in terms of the local potentials such that $\tilde T=\tilde{\bk}\tilde{\mathcal{T}}$ and $\tilde \Omega=\tilde{\bk}\tilde{\omega}$. Therefore we obtain the expression for the brane thickness 
\beq
\tilde{r}_0=\tilde r_+\tilde{\bk}~~,~~\tilde r_+=\frac{(\tilde n-2)}{4\pi \tilde T}~~,
\eeq
and hence the effective action \eqref{eq:freestring1}.

Even though we have derived this action for the case $\tilde p=1$, it is valid for all $\tilde p\ge1$ since we could have added an arbitrary number of flat directions to \eqref{ds:disc} and proceeded in the same way. We also note that, even though the derivation of $\tilde{\mathcal{F}}[\tilde X^{i}]$ was done for flat backgrounds, it still applies to curved backgrounds for which the associated curvature length scales are much larger than $\tilde \omega^{-1}$. Before studying some of the properties of this effective theory, we consider branes of Myers-Perry black holes with an arbitrary number of ultraspins.

\subsection{Effective theory for Myers-Perry branes with several ultraspins}
Instead of considering strings or branes of discs as in \eqref{ds:sdisc}, one can consider as a starting point the geometry of an even-ball instead of a disc \eqref{ds:disc}. Even-ball geometries of dimensionality $2k$ describe Myers-Perry black holes with several ultraspins \cite{Emparan:2009vd} and have induced metric
\beq \label{ds:even}
\textbf{ds}^2=-d\tau^2+\sum_{a=1}^{k}\left(d\rho_a^2+\rho_a^2d\phi_a^2\right)~~.
\eeq
These geometries are also Lorentzian minimal surfaces since they are just embeddings of $\mathbb{R}^{(2k)}$ into $\mathbb{R}^{(D-1)}$ and trivially satisfy the blackfold equations since ${K_{ab}}^{i}=0$. The geometry is rotating with angular velocity $\Omega^a$ on each of the Cartan angles $\phi_a$, hence, according to the blackfold equations \eqref{eq:bfeom2} its boundaries are described by the equation
\beq
\sum_{a=1}^{k}(\Omega^{a})^{2}\rho_{a}^2=1~~.
\eeq 
Proceeding as in the previous section and adding $\tilde p$ flat directions to \eqref{ds:even} one obtains a brane of even-balls. Integrating out the $2k$-dimensional even-ball geometry we obtain an effective description in terms of a $\tilde p$-brane with $k$ particle currents and effective free energy
\beq \label{eff:mp}
\tilde{\mathcal{F}}[\tilde X^{i}]=\frac{\Omega_{(\tilde n+1)}}{16\pi G}\frac{\tilde r_+^{\tilde n -2k}}{\prod_{a=1}^{k}\tilde \Omega_a^2}\int_{\mathcal{B}_{\tilde p}} \sqrt{-\tilde \gamma}~ \tilde{\bk}^{\tilde n}~~.
\eeq
From here, as previously, it is possible to extract the local effective theory which, when restricted to stationary configurations, leads to \eqref{eff:mp}. The local thermodynamic densities valid in the regime $\tilde \omega_a\tilde r_0\ll1,\forall a$ are 
\beq
\!\!\!\tilde P=-\frac{\Omega_{(\tilde n+1)}}{16\pi G} \frac{\tilde r_0^{\tilde n-2k}}{\prod_{a=1}^{k}\tilde{\omega}_a^2}~~,~~\tilde s=\frac{\Omega_{(\tilde n+1)}}{4 G} \frac{\tilde r_0^{\tilde n-2k+1}}{\prod_{a=1}^{k}\tilde{\omega}_a^2}~~,~~\tilde{\mathcal{J}}_{(a)}=\frac{\Omega_{(\tilde n+1)}}{8 \pi G} \frac{\tilde r_0^{\tilde n-2k}}{\tilde{\omega}_{a}\prod_{b=1}^{k}\tilde{\omega}_b^2}~~,
\eeq
with $D=\tilde n+\tilde p +3$ and $\tilde n>2k$ and the local thickness is given by
\beq
\tilde r_0=\frac{(\tilde n-2k)}{4\pi\tilde{\mathcal{T}}}~~.
\eeq
These local thermodynamic densities satisfy the relations
\beq
\tilde{\epsilon}+\tilde P=\tilde{\mathcal{T}}\tilde s+\sum_{a=1}^{k}\tilde{\omega}_a\tilde{\mathcal{J}}_{(a)}~~,~~\tilde{\epsilon}=-(\tilde n+1)\tilde P~~.
\eeq
Hence the effective stress-energy tensor is that of a multi-charged perfect fluid carrying a $k$ number of particle charges. In order to describe stationary configurations one must require $\tilde T=\tilde{\bk} \tilde{\mathcal{T}}$ and $\tilde\Omega_{a}=\tilde{\bk}\tilde{\omega}_a$. From the effective free energy \eqref{eff:mp}, we see that no corrections to the equilibrium condition of black holes appear since the effective action is proportional to $\tilde{\bk}$. In the case of doubly-spinning black rings, this is consistent with the arguments of \cite{Armas:2011uf} for the absence of corrections due to intrinsic spin. The differences, to leading order, rely on their thermodynamic properties.

\subsection{Thermodynamic properties and stability}
In this section we collect the thermodynamic formulae for the conserved charges obtained using this effective theory and comment on the stability properties for this class of branes. As mentioned, the effective theory is characterised by an entropy current $J_s^{a}=\tilde s u^{a}$ and a $k$-number of particle currents $\tilde{\mathcal{J}}^{a}_{(b)}=\tilde{\mathcal{J}}_{(b)}\tilde u^{a}$. Integrating these quantities over the worldvolume as in \cite{Armas:2014rva} leads to the total entropy $S$ and transverse angular momenta $\tilde{J}_{(b)\perp}$ given by
\beq
\!\!\!\!\!\!\tilde S=\frac{\Omega_{(\tilde n+1)}}{4 G} \frac{\tilde r_+^{n-2k+1}}{\prod_{a=1}^{k}\tilde{\Omega}_a^2}\int_{\mathcal{B}_{\tilde p}}dV_{(\tilde p)}~\tilde R_0~\tilde \bk^{\tilde n}~~,~~\tilde{J}_{(b)\perp}=\frac{\Omega_{(\tilde n+1)}}{8\pi G} \frac{\tilde r_+^{n-2k}}{\tilde \Omega_{b}\prod_{a=1}^{k}\tilde{\Omega}_a^2}\int_{\mathcal{B}_{\tilde p}}dV_{(\tilde p)}~\tilde R_0~\tilde \bk^{\tilde n}~~,
\eeq
where we have assumed that the worldvolume time-like Killing vector field is hypersurface orthogonal such that $\sqrt{-\gamma}d^{\tilde p+1}\tilde\sigma=\tilde R_0dV_{(\tilde p)}$. In turn the total mass and angular momenta along worldvolume directions read
\beq
\begin{split}
\tilde M&=\frac{\Omega_{(\tilde n+1)}}{16\pi G} \frac{\tilde r_+^{\tilde n-2k}}{\prod_{a=1}^{k}\tilde{\Omega}_a^2}\int_{\mathcal{B}_{\tilde p}}dV_{(\tilde p)}~\tilde R_0~\tilde \bk^{\tilde n}\left(1+\tilde n\frac{\tilde R_0^2}{\tilde\bk^2}\right)~~,\\
\tilde J_{(b)}&=\frac{\Omega_{(\tilde n+1)}}{16\pi G} \frac{\tilde r_+^{\tilde n-2k}}{\prod_{a=1}^{k}\tilde{\Omega}_a^2}\tilde n \Omega_{b}\int_{\mathcal{B}_{\tilde p}}dV_{(\tilde p)}~\tilde R_0~\tilde R_b^2~\tilde \bk^{\tilde n-2}~~.
\end{split}
\eeq
The form of these expressions is very similar to those obtained for the usual local blackfold effective theory for the branes \eqref{ds:blackp}. In particular we find a similar relation between the free energy \eqref{eff:mp} and the entropy as the one found for charged black branes in \cite{Caldarelli:2010xz}, namely, $\tilde{\mathcal{F}}=\tilde T\tilde S/(\tilde n-2k)$.

\subsubsection*{Gregory-Laflamme and correlated stability}
In the blackfold literature several black branes have been shown to be unstable under long-wavelength perturbations along worldvolume directions \cite{Emparan:2009at, Caldarelli:2010xz, Emparan:2011hg, Gath:2013qya, DiDato:2015dia}. Such instability is known as the Gregory-Laflamme instability. In order to assess if ultraspinning Myers-Perry branes are unstable we evaluate the speed of propagation of sound (longitudinal) and elastic (transverse) waves. For the case of the speed of sound waves $c_s^2$, we use the result of \cite{Gath:2013qya}, where fluids with a particle current were analysed, while for the speed of elastic waves $c_T^2$, the result of \cite{Emparan:2009at} applies,
\beq
c_s^2=\left(\frac{\partial \tilde P}{\partial \tilde \epsilon}\right)_{\frac{\tilde s}{\tilde{\mathcal{J}}}}~~,~~c_T^2=-\frac{\tilde P}{\tilde \epsilon}~~.
\eeq
We find that $c_s^2=-c_T^2=-(\tilde n+1)^{-1}$, which is the same result as for the branes \eqref{ds:blackp} as shown in \cite{Emparan:2009at}. Therefore this class of branes is hydrodynamically unstable and elastically stable, as expected since ultraspinning black holes are known to be unstable. This analysis is expected to change if one considers Myers-Perry branes with arbitrary angular momentum, in which case stable regimes (near extremality) are expected to appear as for fluids carrying particle charge \cite{Caldarelli:2010xz, Gath:2013qya}. In evaluating the speed of sound we have considered only Myers-Perry branes with one single ultraspin, however, we expect the analysis for a single ultraspin to extend to multi-ultraspin cases.

The existence of a Gregory-Laflamme instability is expected to be correlated with thermodynamic instability \cite{Reall:2001ag, Gubser:2000mm, Gubser:2000ec}. In order to see that this is the case for this class of branes we compute the specific heat capacity $C_{\tilde{\mathcal{J}}}$ and inverse isothermal permitivity $c$ for Myers-Perry branes with a single ultraspin. This has to be done in the grand canonical ensemble since the density of transverse angular momentum $\tilde{\mathcal{J}}$ can be redistributed over the worldvolume. We obtain the following results,
\beq
\begin{split}
C_{\tilde{\mathcal{J}}}=\left(\frac{\partial \tilde \epsilon}{\partial \tilde{\mathcal{T}}}\right)_{\tilde{\mathcal{J}}}=-\frac{(\tilde n+1)}{3}\tilde s~~,~~c= \left(\frac{\partial \tilde \omega}{\partial \tilde{\mathcal{J}}}\right)_{\tilde{\mathcal{T}}}=-\frac{\tilde\omega}{3\tilde{\mathcal{J}}}~~.
\end{split}
\eeq
We see that $C_{\tilde{\mathcal{J}}}<0$ and $c<0$ for all values of the thermodynamic variables. Note that $\tilde\omega/\tilde{\mathcal{J}}\propto \tilde\omega^4$ and hence is independent of the sign of $\tilde \omega$. Therefore, we conclude that for this class of branes, the Gregory-Laflamme instability is correlated with thermodynamic instability. This behaviour is qualitatively different from the analysis for fluids with particle charge carried out in \cite{DiDato:2015dia}.

\subsection{Higher-order transport} \label{sec:hight}
As mentioned in Sec.~\ref{sec:int}, the procedure of obtaining the effective free energy functional holds to higher orders in the derivative expansion and hence it can be used to obtain higher-order transport coefficients of other effective theories from the original ones. However, this procedure works in both directions and hence, if one knows the transport coefficients of the resulting effective theory, one can obtain the transport coefficients of the original theory. In this section we give an example for each of these situations, focussing on the hydrodynamic transport coefficient $\upsilon_1$ and corresponding $\tilde \upsilon_1$, while in Sec.~\ref{sec:blacktor} we obtain one contribution to the Young modulus of ultraspinning Myers-Perry branes involving the transport coefficients $\lambda_i$. We will in particular show how information about the leading order effective theory of Myers-Perry branes can give us access to second order transport coefficients characterising the branes \eqref{ds:blackp}.

\subsubsection*{The transport coefficient $\upsilon_1$}
We first consider the higher-order corrections that arise for the simplest configuration \eqref{ds:sdisc}. Since it is a flat worldvolume embedded into flat space-time with vanishing extrinsic curvature tensor, the only non-zero invariant that contributes to the second order effective free energy \eqref{eq:free} is the invariant $|\mathfrak{a}^{c}\mathfrak{a}_{c}|$. Explicit computation for the geometry \eqref{ds:sdisc} leads to
\beq
\mathfrak{a}^{c}\mathfrak{a}_{c}=\frac{\rho^2\Omega^{4}}{\bk^4}~~.
\eeq
The effective free energy functional \eqref{eq:free} for this configurations is thus
\beq \label{corr2}
\!\!\!\mathcal{F}=-\int_{\mathcal{W}_{p+1}}dV_{(p)}\left(P+\upsilon_{1}\mathfrak{a}^{c}\mathfrak{a}_{c}\right)=-\frac{\Omega_{(\tilde n+1)}}{16\pi G}\frac{\tilde r_+^{\tilde n -2}}{\tilde\Omega^2}\int_{\tilde{\mathcal{B}}_{\tilde p}} \sqrt{-\tilde \gamma}~ \tilde{\bk}^{\tilde n}\left(-1+\frac{2\alpha}{\tilde n-2} \tilde r_+^{2}\tilde\Omega^2\right)~~,
\eeq
where we have written $\upsilon_1$ as
\beq
\upsilon_1=\frac{\Omega_{(n+1)}}{16\pi G}r_0^{n+2}\alpha~~,
\eeq
for some constant $\alpha$. Since all the second order invariants written in \eqref{eq:freet} vanish for the resulting effective geometry \eqref{eq:effgeo}, we interpret the resulting correction in \eqref{corr2} of order $(r_+\tilde \Omega)^2$ as a correction to the local effective pressure \eqref{p:mp}, that is, a correction of order $\ell^2$ to the underlying local effective theory with scale parameter $\ell=\tilde r_0\tilde\omega$. Writing it in terms of local quantities we obtain
\beq\label{pmp1}
\tilde P=-\frac{\Omega_{(\tilde n+1)}}{16\pi G} \frac{\tilde r_0^{\tilde n-2}}{\tilde{\omega}^2}\left(1-\frac{2\alpha}{\tilde n-2}\tilde r_0^2\tilde\omega^2\right)~~.
\eeq
We see that corrections proportional to $\mathfrak{a}^{c}\mathfrak{a}_{c}$ in the effective theory \eqref{eq:free} correspond to moving further way from the ultraspinning limit of Myers-Perry branes. Since in \cite{Armas:2011uf} the effective pressure $P_{MP}$ of Myers-Perry branes was obtained, we can expand it for large rotation parameter.\footnote{In the notation of \cite{Armas:2011uf}, this corresponds to expanding the pressure for large rotation parameter $b$.} We obtain
\beq \label{pmp2}
P_{MP}=-\frac{\Omega_{(\tilde n+1)}}{16\pi G} \frac{\tilde r_0^{\tilde n-2}}{\tilde{\omega}^2}\left(1-\tilde r_0^2\tilde\omega^2+\mathcal{O}\left((\tilde r_0\tilde\omega)^4\right)\right)~~.
\eeq
Comparing \eqref{pmp2} with \eqref{pmp1} leads to $\alpha=(\tilde n-2)/2$, which when written in terms of $n$ leads to the result claimed in \eqref{up1}. This result is valid for any configuration for which the fluid vorticity $\omega_{ab}$ vanishes.

\subsubsection*{The transport coefficient $\tilde \upsilon_1$}
Given that we have determined $\upsilon_1$ we can use it in order to determine the corresponding transport coefficient for ultraspinning Myers-Perry branes. Consider the flat embedding geometry, consisting of adding a two-plane to \eqref{ds:sdisc}
\beq
\textbf{ds}^2=-d\tau^2+d\rho_1^2+\rho_1^2d\phi_1^2+d\rho_2^2+\rho_2^2d\phi_2^2+dz^2~~,
\eeq
where the Killing vector field takes the form
\beq
\textbf{k}^a\partial_a=\partial_\tau+\Omega_1\partial_{\phi_1}+\Omega_2\partial_{\phi_2}+\beta\partial_z~~.
\eeq
For this geometry all the invariants in \eqref{eq:free} vanish except for the one proportional to the square of the acceleration, which reads
\beq \label{count4}
\mathfrak{a}^{c}\mathfrak{a}_{c}=\frac{\rho_1^2\Omega_1^{4}}{\bk^4}+\frac{\rho_2^2\Omega_2^{4}}{\bk^4}~~.
\eeq
We will integrate only over the disc spanned by $(\rho_1,\phi_1)$ with boundary $\rho_{1+}=\sqrt{1-\Omega_2^2\rho_2^2}/\Omega_1$. In this way all the invariants in \eqref{eq:freet} for the resulting effective geometry vanish except for the one proportional to the acceleration $\tilde{\mathfrak{a}}^{c}\tilde{\mathfrak{a}}_{c}$, which reads
\beq
\tilde{\mathfrak{a}}^{c}\tilde{\mathfrak{a}}_{c}=\frac{\rho_2^2\Omega_2^{4}}{\tilde \bk^4}~~.
\eeq
The integration over the contribution \eqref{count4} leads to one part which gives rise to the correction to the effective pressure as in \eqref{pmp1}, while the remaining term leads to a contribution to the effective free energy \eqref{eq:freet} of the form
\beq
-\int_{\tilde{\mathcal{B}}_{\tilde p}}\sqrt{-\tilde\gamma}\frac{\Omega_{(\tilde n+1)}}{16\pi G}\frac{\tilde n}{2}\tilde r_0^{\tilde n+2}\tilde{\mathfrak{a}}^{c}\tilde{\mathfrak{a}}_{c}~~.
\eeq
Therefore we obtain the transport coefficient $\tilde \upsilon_1$ for Myers-Perry branes in the ultraspinning regime
\beq
\tilde \upsilon_1=\frac{\Omega_{(\tilde n+1)}}{16\pi G}\frac{\tilde n}{2}\tilde r_0^{\tilde n+2}~~,~~\tilde n>4~~.
\eeq

%%%%%%%%%%%%%%%%%%%%%%%%%%%%%%%%%%%%%%
%%%%%%%%%%%%%%%%%%%%%%%%%%%%%%%%%%%%%% 
%%%%%%%%%%%%%%%%%%%%%%%%%%%%%%%%%%%%%%

\section{Effective theory for helicoidal black branes} \label{sec:effhelicoids}
In this section we construct another example of an effective theory for black branes based on the minimal surface embedding of the helicoid and one of its higher-dimensional versions found in \cite{Armas:2015kra}. The resulting effective theory is non-trivial and the end result of the process of integrating out part of the worldvolume geometry gives rise to a local brane metric which is not known. Before we proceed and derive the effective action for these branes we review and present a careful study of the geometry of a black helicoid, which is a novel solution of the blackfold equations \eqref{eq:bfeom2} found in \cite{Armas:2015kra}.

%%%%%%%%%%%%%%%%%%%%%%%%%%%%%%%%%%%%%% 
%%%%%%%%%%%%%%%%%%%%%%%%%%%%%%%%%%%%%%
%%%%%%%%%%%%%%%%%%%%%%%%%%%%%%%%%%%%%% 
\subsection{Black helicoids} \label{sec:helicoids}
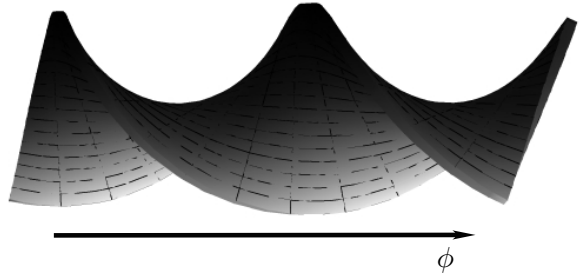
\begin{figure}[h!] 
\centering
  \includegraphics[width=0.6\linewidth]{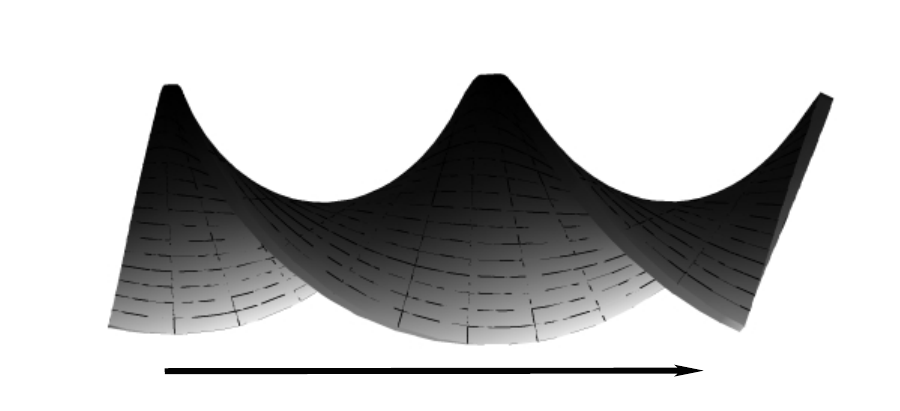}
  \begin{picture}(0,0)(0,0)
\put(-80,0){ $ \phi $}
\end{picture} 
\caption{Embedding of the rotating black helicoid in $\mathbb{R}^{3}$ with $\lambda=a=\Omega=1$, depicted in the interval $-3\le\phi\le3$. } \label{fig:helicoid}
\end{figure}

Helicoid geometries are minimal surfaces in $\mathbb{R}^{3}$, which trivially solve the blackfold equations \eqref{eq:bfeom2} in flat space-time, as shown in \cite{Armas:2015kra}. These geometries are embedded in flat space-time with coordinates $(t,x_i)$ according to the embedding map
\beq \label{emb:helic}
t=\tau~~,~~X^{1}(\rho,\phi)=\rho\cos(a\phi)~~,~~X^{2}(\rho,\phi)=\rho\sin(a\phi)~~,~~X^{3}(\rho,\phi)=\lambda \phi~~,
\eeq
and $X^{i}=0~,~i=4,...,D-1$, where the coordinates lie within the range $-\infty<\rho,\phi<\infty$. The ratio $\lambda/a$ is the pitch of the helicoid and if $\lambda\ne0$, the coordinate $\phi$ can always be rescaled such that $a$ can be set to 1.  Without loss of generality, we take $\lambda\ge0$ and $a>0$. The induced metric then takes the form
\beq \label{ds:helicoid}
\textbf{ds}^2=-d\tau^2+d\rho^2+(\lambda^2+a^2\rho^2)d\phi^2~~.
\eeq
In order to make one of the directions compact, the helicoid is boosted along the $\phi$ direction with boost velocity $\Omega$ such that
\beq \label{kill:helicoid}
\textbf{k}^{a}\partial_a=\partial_\tau+\Omega\partial_\phi~~,~~\textbf{k}^2=1-\Omega^2(\lambda^2+a^2\rho^2)~~.
\eeq
From the point of view of the background space-time, this corresponds to a Killing vector field $k^{\mu}$ which takes the form
\beq \label{eq:killhell}
k^{\mu}\partial_\mu=\partial_t+a\Omega\left(x_1\partial_{x_{2}}-x_2\partial_{x_1}\right) +\lambda\Omega \partial_{x_3}~~,
\eeq
and hence it is rotating in the $(x_1,x_2)$ plane with angular velocity $a\Omega$ and it is boosted along the $x_3$ direction with boost velocity $\lambda\Omega$.
From Eq.~\eqref{kill:helicoid} and the blackfold equations \eqref{eq:bfeom2} the geometry has boundaries when $\bk=0$, which appear at $\rho_{\pm}$,
\beq \label{eq:reqh}
\rho_{\pm}=\pm\frac{\sqrt{1-\Omega^2\lambda^2}}{a\Omega}~~,~~\Omega\lambda<1~~.
\eeq
The boundary makes the helicoid compact in the $\rho$ direction and give rise to black hole horizons with black string topology $\mathbb{R}\times \mathbb{S}^{(D-3)}$ in $D\ge6$. They can be thought as the membrane analog of the helical strings found in \cite{Emparan:2009vd}. The fact that these geometries have string topology suggests that they can be bent into a helicoidal ring, in the same way that helical strings can be bent into helical rings \cite{Emparan:2009vd}. As we will show in Sec.~\ref{sec:helicoidalrings}, this is indeed the case.\footnote{We thank Roberto Emparan for suggesting this possibility to us.} The size of the transverse sphere $r_0(\rho)$ is given by
\beq
r_0(\rho)=\frac{n}{4\pi T}\sqrt{1-\Omega^2(\lambda^2+a^2\rho^2)}~~,
\eeq
and is maximal at the origin $\rho=0$ and vanishes at the boundaries $\rho_\pm$. This geometry is depicted in Fig.~\ref{fig:helicoid} and, as shown in \cite{Armas:2015kra}, is valid in the regime $r_0\ll \lambda/(\sqrt{2}a)$, $r_+\ll 1/(a\Omega)$ and $r_+\ll \rho_+$.

It is important to mention that the black helicoid is not a helical string with non-zero transverse angular momentum, that is, it is not the geometry of a helical string parametrised by a coordinate $\phi$ in which $\phi=constant$ slices have the geometry of a disc. This would be a 3-brane with the worldvolume geometry of a non-compact helical cylinder. In other words, these geometries are not described by an action of the form \eqref{eff:mp}.\footnote{Note that certain restrictions exist if one wants to add transverse angular momentum to a helical string by wrapping Myers-Perry strings \cite{Emparan:2009vd}.}

\subsubsection*{The free energy and the Myers-Perry limit}

The free energy of black helicoids can be obtained by evaluating \eqref{eq:free} to leading order, yielding
 \beq \label{eq:freehelicoidflat}
 \begin{split}
 \mathcal{F}&=\frac{\Omega_{(n+1)}}{16\pi G}r_+^{n}\int d\phi\int_{\rho_-}^{\rho_+} d\rho\sqrt{\lambda^2+a^2\rho^2}\left(1-\Omega^2(\lambda^2+a^2\rho^2)\right)^{\frac{n}{2}}\\
&= \frac{V_{(n+2)}}{16 \pi G}\frac{r_+^{n}}{a\Omega}\int d\phi \lambda \left(1-\lambda ^2 \Omega
   ^2\right)^{\frac{n+1}{2}} \,
   _2\tilde{F}_1\left(-\frac{1}{2},\frac{1}{2};\frac{n+3}{2};1-\frac{1}{\lambda ^2 \Omega
   ^2}\right)~~,
 \end{split}
 \eeq
where we have defined $V_{(n+2)}=2\pi^{\frac{n+3}{2}}$. The free energy is positive for all $n$ and, since the geometry is non-compact in the $\phi$ direction, it is obviously infinite, as expected since it has the topology of a black string.

The geometry \eqref{ds:helicoid} as well as the free energy \eqref{eq:freehelicoidflat} admits a very well defined limit $\lambda\to0$ in which one recovers the disc. As explained in \cite{Armas:2015kra}, the range of $\rho$ lies in between $\rho_-\le\rho\le\rho_+$ instead of $0\le\rho\le\rho_+$, therefore, when taking the limit $\lambda\to0$ we simultaneously rescale $ \mathcal{F}\to(1/2) \mathcal{F}$. Formally, when taking the limit $\lambda\to0$ in the free energy \eqref{eq:freehelicoidflat} we keep $a$ fixed, make the $\phi$-coordinate periodic with period $2\pi/a$ and integrate $\phi$ in the interval $0\le\phi\le2\pi/a$. The free energy \eqref{eq:freehelicoidflat} then reduces to the result for the disc, when setting $a=1$ and rescaling $ \mathcal{F}\to(1/2) \mathcal{F}$, given by \cite{Emparan:2009vd},
\beq \label{free:mp22}
\mathcal{F}=\frac{\Omega_{(n+1)}}{8G}\frac{r_+^{n}}{(n+2)\Omega^2}~~.
\eeq
It is important to stress that the existence of this limit is non-trivial since it does not change the dimensionality of the worldvolume neither of the space-time. It is not equivalent, for example, to taking the string of discs \eqref{ds:sdisc} and getting rid of the extra dimension $z$ (for example by scaling $z\to\lambda z$ and then taking $\lambda\to0$). This non-trivial equivalence with the geometry and thermodynamics of the disc in the limit $\lambda\to0$ suggests that the family of singly-spinning Myers-Perry black holes can be obtained from the family of black helicoids, at least in the ultraspinning limit, in which the topology changes according to $\mathbb{R}\times \mathbb{S}^{D-3}\to\mathbb{S}^{D-2}$ \cite{Armas:2015kra}.

\subsubsection*{Thermodynamics}
The thermodynamic properties of these geometries can be obtained from the free energy \eqref{eq:freehelicoidflat} by taking the appropriate derivatives \eqref{thermo1}-\eqref{thermo2}. We find the following expressions for the mass $M$, angular momentum $J_{\perp}$ in the $(x_1,x_2)$ plane and linear momentum $\mathcal{P}$ of the helicoid in the $x_3$-direction,
 \beq \label{eq:masshell}
 \begin{split}
 M=\frac{V_{(n+2)}}{32 \pi G}\frac{r_+^{n}}{a\lambda\Omega^3}\!\!\int\!\! d\phi&\left(1\!-\!\lambda ^2 \Omega^2\right)^{\frac{n-1}{2}}\Big( 2\lambda ^2 \Omega ^2 \left(n\!+\!2-\!\lambda ^2 \Omega ^2\right)
   \, \!\!_2\tilde{F}_1\left(-\frac{1}{2},\frac{1}{2};\frac{n+3}{2};1\!-\!\frac{1}{\lambda ^2 \Omega
   ^2}\right)~~\\
&+ \left(1-\lambda ^2 \Omega ^2\right) \,
   _2\tilde{F}_1\left(\frac{1}{2},\frac{3}{2};\frac{n+5}{2};1-\frac{1}{\lambda ^2 \Omega
   ^2}\right) \Big) ~~,
   \end{split}
 \eeq
 \beq
 \begin{split} \label{eq:anghell1}
J_{\perp}=\frac{V_{(n+2)}}{16 \pi G}\frac{r_+^{n}}{(a\Omega)^2}\int d\phi \lambda \left(1-\lambda ^2 \Omega
   ^2\right)^{\frac{n+1}{2}} \,
   _2\tilde{F}_1\left(-\frac{1}{2},\frac{1}{2};\frac{n+3}{2};1-\frac{1}{\lambda ^2 \Omega
   ^2}\right)~~,
   \end{split}
 \eeq
\beq
 \begin{split} \label{eq:anghell2}
 \mathcal{P}=\frac{V_{(n+2)}}{32\pi  G}\frac{ r_+^{n}}{a\lambda^3\Omega^4}\int d\phi&\left(1\!-\!\lambda ^2 \Omega^2\right)^{\frac{n-1}{2}} \Big(\left(1-\lambda ^2 \Omega ^2\right) \,
   _2\tilde{F}_1\left(\frac{1}{2},\frac{3}{2};\frac{n+5}{2};1-\frac{1}{\lambda ^2 \Omega
   ^2}\right)\\
   &+2(\lambda\Omega)^{4}(n+1) \,
   _2\tilde{F}_1\left(-\frac{1}{2},\frac{1}{2};\frac{n+3}{2};1\!-\!\frac{1}{\lambda ^2 \Omega
   ^2}\right)\Big)~~,
   \end{split}
 \eeq
where we have used that $J_{\perp}=\partial \mathcal{F}/\partial (a\Omega)$ and $\mathcal{P}=\partial \mathcal{F}/\partial (\lambda\Omega)$. The entropy is given by $S=(n/T)\mathcal{F}$. These expressions satisfy the relation $\mathcal{F}=M-TS-a\Omega J_{\perp}-\lambda\Omega\mathcal{P}$. In the limit $\lambda\to0$ as explained above, they also reduce to those of the disc obtained in \cite{Emparan:2009vd}.

These geometries, because they are non-compact in the $\phi$ direction, do not in general satisfy the Smarr relation \eqref{eq:smarr} in global asymptotically flat space-time. Therefore they have a non-zero tension \eqref{eq:tension} given by
\beq\label{eq:tensionhelicoid}
 \begin{split}
 \boldsymbol{\mathcal{T}}=-\frac{V_{(n+2)}}{32 \pi G}&\frac{r_+^{n} \left(1\!-\!\lambda ^2 \Omega^2\right)^{\frac{n-1}{2}}}{a\lambda\Omega^3}\int d\phi\Big( (1-\lambda^2\Omega^2)\, \,_2\tilde{F}_1\left(\frac{1}{2},\frac{3}{2};\frac{n+5}{2};1-\frac{1}{\lambda ^2
   \Omega ^2}\right)  \\
  &-2\lambda^2 \Omega ^2(1-(n+2)\lambda^2\Omega^2) \,
   _2\tilde{F}_1\left(-\frac{1}{2},\frac{1}{2};\frac{n+3}{2};1-\frac{1}{\lambda ^2 \Omega
   ^2}\right)\Big)~~.
   \end{split}
 \eeq
In the limit $\lambda\to0$ this becomes the disc and hence $\boldsymbol{\mathcal{T}}\to0$ and the Smarr relation \eqref{eq:smarr} for compact black holes in flat space-time is satisfied. We note, however, that there is a specific value of $\Omega$ for each $n$ that leads to a vanishing integrand in the tension \eqref{eq:tensionhelicoid}. This specific value is the one required to balance helicoidal black rings, as we will see in Sec.~\ref{sec:helicoidalrings}.

\subsubsection*{Saturating the rigidity theorem}
Black helicoids can be seen as the membrane generalisation of the helical strings found in \cite{Emparan:2009vd}. In particular, their worldvolume geometry both preserve one $U(1)$ family of spatial isometries whose orbits are not closed. In this case, this is given by the spatial part of the Killing vector \eqref{eq:killhell} associated with the total conserved momentum $(aJ_{\perp}+\lambda \mathcal{P})$. Therefore, some of the considerations of \cite{Emparan:2009vd} for helical strings also apply to the case of black helicoids. Namely, if the helicoid geometry winds all the $[(D-1)/2]$ planes of the background space-time then the resulting configuration will break the spherical symmetries of the transverse space and can at most preserve one family of spatial isometries. Hence, it constitutes another example of a black hole which saturates the rigidity theorem \cite{Hollands:2006rj}. The form of the embedding in \eqref{emb:helic} is only winding around one plane but it is trivial to extend the embedding map to the case where it is winding an arbitrary number of planes. This implies that the helicoidal black rings constructed in Sec.~\ref{sec:helicoidalrings} can also saturate the rigidity theorem.

%%%%%%%%%%%%%%%%%%%%%%%%%%%%%%%%%%%%%
\subsection{Black helicoid $p$-branes}  \label{sec:higherhelicoid}
In this section we review the higher-dimensional analogue of the black helicoids of the previous section as in \cite{Armas:2015kra}. These helicoids, also known as the Barbosa-Dajczer-Jorge helicoids \cite{Jorge:1984}, can be explicitly embedded into a subspace $\mathbb{R}^{2k+1}$, where $k\ge1$ is an integer, of $\mathbb{R}^{(D-1)}$ according to \cite{LeeLee:2014}
 \beq \label{emb:higherhell}
 \begin{split}
 X^{q}(\rho_q,\phi) &= \rho_q \cos (a_q \phi)~\text{if $q$ is odd and $1\le q\le 2k$}~~,\\
 X^{q}(\rho_q,\phi)&= \rho_{q-1} \sin (a_{q-1} \phi)~\text{if $q$ is even and $1\le q\le 2k$}~~,\\
X^{q}(\rho_q,\phi)&=\lambda ~\phi~\text{if $q=2k+1$}~~,
 \end{split}
 \eeq
 and $t=\tau~,X^{i}=0~,~i=2k+2,...,D-1$. Here $a_q,\lambda$ are constants which without loss of generality we require to satisfy $a_q>0$ and $\lambda>0$. Note that $k$ and $p$ are related such that $p=k+1$. The coordinates lie within the range $-\infty<\rho_q,\phi<\infty$. For $k=1$ we obtain the two dimensional helicoid of the previous section. In general we can rescale $\phi$ such that $\phi\to \lambda^{-1}\phi$ and $a_q \to  \lambda^{-1} a_q$ and hence set $\lambda=1$. The case $\lambda=0$ represents a minimal cone, which has a conical singularity at the origin and therefore we do not consider it. The induced metric on the worldvolume takes the form
\beq \label{ds:helicalp}
 \textbf{ds}^2=-d\tau^2+\sum_{a=1}^{k}d\rho_{a}^2+\left(\lambda^2+\sum_{a=1}^{k}a_a^2\rho_{a}^2\right)d\phi^2~~,
 \eeq
and it has boost velocity $\Omega$ along the $\phi$-direction such that $\textbf{k}^{a}\partial_a=\partial_\tau+\Omega\partial_\phi$. This maps onto the vector field in the background space-time 
\beq \label{kill:hell}
k^{\mu}\partial_\mu=\partial_t +\Omega\sum_{a=1}^{k}a_{a}\left(x_{2a-1}\partial_{x_{2a}}-x_{2a}\partial_{x_{2a-1}}\right)+\lambda\Omega\partial_{x_{2k+1}}~~.
\eeq
According to the blackfold equations \eqref{eq:bfeom2}, the boundaries of the worldvolume geometry are defined by the equation
\beq \label{line}
\sum_{a=1}^{k}a_a^2\rho_{a}^2=\frac{1-\Omega^2R^2}{\Omega^2}~~.
\eeq
These configurations in $D\ge6$ with $D-p-3\ge1$ can be though of as $p$-brane version of the helical strings found in \cite{Emparan:2009vd}. They again have the topology of a black string, that is, $\mathbb{R}\times\mathbb{S}^{(D-3)}$ and are valid in the regime $r_0\ll \frac{\lambda}{\sqrt{\sum_{a=1}^{N}a_{a}^2}}$, $r_+(a_a\Omega)\ll1$ and $r_+~\ll \rho_{a}^{+}$ \cite{Armas:2015kra}. Their free energy can be easily evaluated using \eqref{eq:free}
\beq \label{free:helicp}
\!\!\!\!\!\!\!\!\mathcal{F}=\frac{V_{(n+p)}}{16\pi G}\frac{\lambda r_+^{n}}{\Omega^{p-1}\prod_{a=1}^{p-1}a_a}\int d\phi \left(1\!-\!\lambda ^2 \Omega
   ^2\right)^{\frac{n+p-1}{2}} \!\!
   _2\tilde{F}_1\left(-\frac{1}{2},\!\frac{(p\!-\!1)}{2};\!\frac{n\!+\!p\!+\!1}{2};\!1\!-\!\frac{1}{\lambda ^2 \Omega
   ^2}\right),
\eeq
where $V_{(n+p)}=2\pi^{\frac{n+3}{2}}$ if $p=2$ and $V_{(n+p)}=4\pi^{\frac{n+p+1}{2}}$ if $p>2$. From here as in the $p=2$ case of the previous section, we can obtain all thermodynamic properties as we will show explicitly in the next section.

\subsection{Effective theory for helicoidal black $p$-branes}  
The helicoid $p$-branes of the previous section have topology $\mathbb{R}\times\mathbb{S}^{(D-3)}$ and therefore we refer to them as \emph{helicoidal black strings}. The induced spatial geometry \eqref{ds:helicalp} is of the form $\mathbb{E}_{M}^{(p)}=\mathbb{R}\times \mathbb{I}^{(p-1)}$, where $\mathbb{I}$ denotes the topology associated with each finite interval $\rho_{q}$. Therefore, the worldvolume has several distinct length scales $\ell_{\hat a}$: the ones associated with the $(p-1)$-worldvolume directions of the helicoidal brane and the one associated with the infinitely extended $\phi$-coordinate. The dynamics of the fluid decouples on the two spaces $\mathbb{R}$ and $\mathbb{I}^{(p-1)}$, hence we can integrate out the $(p-1)$ spatial sections of the worldvolume geometry \eqref{ds:helicalp} and obtain the effective geometry of a boosted helical string
\beq \label{effectivestring}
\tilde{\textbf{ds}}^2=-d\tau^2+\lambda^2d\phi^2~~,~~\tilde{\bk}^{a}\partial_a=\partial_\tau+\Omega\partial_\phi~~,~~\tilde{\bk}^2=1-\lambda^2\Omega^2~~.
\eeq
This results in an effective theory of a boosted string with a nontrivial boost vector $\tilde v^{a}$.  Since we have integrated out $k=p-1$ spatial dimensions, we parametrise the number of space-time dimensions as $D=\tilde n+\tilde p+3$ where $\tilde p=1$ and $n=\tilde n-k$ with $\tilde n>k$. With this we then write the effective action for helicoidal black strings \eqref{free:helicp} as
\beq \label{eff:2}
\tilde{\mathcal{F}}[\tilde X^{i}]=\frac{V_{(\tilde n+1)}}{16\pi G}\frac{\tilde r_+^{\tilde n-k}}{\prod_{a=1}^{k}\tilde\Omega_a}\int_{\mathcal{B}_{\tilde p}}\sqrt{-\tilde\gamma}~\tilde{\bk}^{\tilde n}\thinspace _2\tilde{F}_1\left(-\frac{1}{2},\!\frac{k}{2};\!\frac{\tilde n+2}{2};-\frac{\tilde\bk^2}{\tilde{\textbf{v}}^2}\right)~~,
\eeq
where we have defined $\tilde r_+=(\tilde n-k)/4\pi T$, $\tilde\Omega=a_a\Omega$ and also introduced the vector $\tilde{\textbf{v}}^a\partial_a=\Omega \partial_\phi$ with $\tilde{\textbf{v}}=|\gamma_{ab}\tilde{\textbf{v}}^a\tilde{\textbf{v}}^b|^{1/2}$ being its modulus. As we shall see below, the effective action \eqref{eff:2} also holds for a particular case of helicoidal branes. In the case $\tilde p=1$ we note that we can also write $\tilde{\textbf{v}}^2=1-\tilde \bk^2$. The vector $\tilde{\textbf{v}}^a$ is an example of the type of boost vectors, discussed in Sec.~\ref{sec:int}, which can appear in effective theories after integrating out specific degrees of freedom.

\subsubsection*{A proposal for the underlying local effective theory}

The appearance of the modulus of the vector $\tilde{\textbf{v}}^a\partial_a$ in the effective action suggests that the underlying theory does not have a rest frame as the limit $\tilde{\textbf{v}}\to0$ leads to a divergent free energy. We now proceed and make a proposal for the local effective theory which leads to \eqref{eff:2} when restricted to its stationary sector. Using several clues, such as the stress-energy tensor that can be derived from \eqref{eff:2}, and a bit of guesswork leads us to consider the following local Gibbs free energy
\beq \label{localgibbs}
\mathcal{\tilde{G}}=\frac{V_{(\tilde n+1)}}{16\pi G}\frac{\tilde r_0^{\tilde n-k}}{\prod_{a=1}^{k}\tilde\omega_a}f(-\frac{1}{2},k,\tilde n,\Xi)~~,
\eeq
valid in the regime $\tilde \omega_a\tilde r_0\ll1$, or alternatively $\tilde r_0\ll\ell_{\hat a}$ with $\ell_{\hat a}=\tilde\omega_{a}^{-1}$, where we have defined the local brane thickness
\beq \label{localvar}
\tilde r_0=\frac{(\tilde n -k)}{4\pi \tilde{\mathcal{T}}}
\eeq
and also defined for convenience
\beq
f(-\frac{1}{2},k,\tilde n,\Xi)=\thinspace _2\tilde{F}_1\left(-\frac{1}{2},\!\frac{k}{2};\!\frac{\tilde n\!+\!2}{2};-\frac{1}{\Xi^2}\right)~~.
\eeq
We assume that every scalar characterising the fluid can be expressed as a function of $k+2$ fluid dynamical variables, namely, the thermodynamic potentials $\tilde{\mathcal{T}}$, $\tilde\omega_{a}$ and $\Xi$. Using a local version of \eqref{thermo2} we extract from \eqref{localgibbs} the corresponding conjugate thermodynamic variables, namely, the entropy density $\tilde s$, the particle densities of transverse angular momentum $\tilde{\mathcal{J}}_{(a)}$ and a particle current density of linear momentum $\tilde{\mathcal{P}}_{\Xi}$, for which its interpretation will be clearer below. These are given by,
\beq
\!\!\!\!\tilde s=\frac{(\tilde n-k)}{\tilde{\mathcal{T}}}\mathcal{\tilde{G}}~~,~~\tilde{\mathcal{J}}_{(a)}=\frac{1}{\tilde{\omega}_a}\mathcal{\tilde{G}}~~,~~\tilde{\mathcal{P}}_{\Xi}=\frac{k}{2\Xi^3}\frac{f(\frac{1}{2},k+2,\tilde n+2,\Xi)}{f(-\frac{1}{2},k,\tilde n,\Xi)}\mathcal{\tilde{G}}~~,
\eeq
where the momentum density is defined as $\tilde{\mathcal{P}}_{\Xi}=-\partial\mathcal{\tilde{G}}/\partial\Xi$. The free energy \eqref{localgibbs} should be interpreted as the free energy characterising a black brane with $k$ transverse angular momenta $\tilde{\mathcal{J}}_{(a)}$ and another conserved charge $\tilde{\mathcal{P}}_{\Xi}$, which cannot be boosted away via a local Lorentz transformation.

From \eqref{eff:2} we can obtain the stress-energy tensor $\tilde T^{ab}$ for stationary configurations. Based on that we consider the following form for the stress-energy tensor, written in terms of local quantities,
\beq \label{st:eff}
\tilde T^{ab}=\tilde P \tilde{\gamma}^{ab}+\left(\tilde{\mathcal{T}}\tilde s+\sum_{a=1}^{k}\tilde \omega_{a}\tilde{\mathcal{J}}_{(a)}+\Xi \tilde{\mathcal{P}}_{\Xi}\right)\tilde u^{a}\tilde u^{b}-\Xi \tilde{\mathcal{P}}_{\Xi}\tilde{v}^{a}\tilde{v}^{b}~~,
\eeq
where we have defined the unit normalised spatial vector $\tilde v^{a}$ such that $\tilde u^{a}\tilde v_{a}=\Xi$ and used that $\tilde P=-\mathcal{\tilde{G}}$. The vector $\tilde u^{a}$ is normalised such that $\tilde u^{a}\tilde u_{a}=-1$ while the vector $\tilde v^{a}$ is normalised such that $\tilde v^{a}\tilde v_a=1$. We further assume that $\tilde v^{a}$ only has a non-vanishing component in the spatial helicoidal direction which we denote by $\phi$. From these definitions we see that $\Xi$ parametrises the necessary non-zero boost along the helicoidal direction and hence its conjugate variable $\tilde{\mathcal{P}}_{\Xi}$ can be interpreted as a momentum density along that direction carried by the non-vanishing vector $\tilde v^{a}$. The stress-energy tensor \eqref{st:eff} is thus written in a (non-standard) boosted frame due to the fact that the limit $\tilde{\textbf{v}}\to0$ in the effective action \eqref{eff:2} leads to a divergent result. 

Counting the total number of fluid dynamical variables, we see that we have $\tilde p-1$ independent components of $\tilde u^{a}$ (the normalisation condition $\tilde u^{a}\tilde u_{a}=-1$ and condition $\tilde u^{a}\tilde v_{a}=\Xi$ eliminate two possibly independent components), and $k+2$ thermodynamic potentials associated with $\tilde{\mathcal{T}}$, $\tilde\omega_{a}$ and $\Xi$. We note that there are no independent components of $\tilde v^{a}$ since it is completely fixed by its normalisation. Therefore, there is a total of $\tilde p+k+1$ fluid dynamical variables. Their evolution is determined by the conservation of the stress-energy tensor and conservation of transverse angular momentum currents,
\beq
\nabla_a \tilde T^{ab}=0~~,~~\nabla_{a}\left(\tilde{\mathcal{J}}_{(a)} \tilde u^{a}\right)=0~~.
\eeq
Hence there is a total of $\tilde p+k+1$ independent equations which completely specify the evolution of the system. The requirement for the conservation of stress-energy tensor comes from the probe equations \eqref{eq:bfeom} while the requirement for the conservation of transverse angular momenta can be derived for any fluid rotating in transverse directions to the worldvolume \cite{Armas:2013hsa, Armas:2014rva}. In terms of symmetries, the fluid is characterised by a set of $k$ $U(1)$ symmetries associated with the conserved transverse angular momenta and a maximum of $\tilde p+1$ worldvolume translations associated with mass and momentum along worldvolume directions. The currents which correspond to these translations are simply given by $\tilde T^{ab}\tilde{\textbf{k}}_{b}$, for any worldvolume Killing vector field $\tilde{\textbf{k}}^{b}$, and are conserved due to the symmetry of the stress-energy tensor. Some of these $k$ transverse rotations and $\tilde p+1$ worldvolume translations can be broken globally depending on the fluid configuration.

The remaining thermodynamic properties of the fluid, such as the effective energy density $\tilde \epsilon$, can be obtained by performing the Legendre transform in the local Gibbs free energy density \eqref{localgibbs}, giving rise to
\beq
\tilde \epsilon=-(\tilde n+1)\tilde P +\sum_{a=1}^{k}\tilde \omega_{a}\tilde{\mathcal{J}}_{(a)}+\Xi \tilde{\mathcal{P}}_{\Xi}~~,
\eeq
and hence to the Euler-Gibbs-Duhem relation
\beq \label{gibbs1}
(\tilde \epsilon+\tilde P)=\tilde{\mathcal{T}}\tilde s+\sum_{a=1}^{k}\tilde \omega_{a}\tilde{\mathcal{J}}_{(a)}+\Xi \tilde{\mathcal{P}}_{\Xi}~~.
\eeq 
We note that $\tilde u^{a}$ is not an eigenvector of the stress-energy tensor \eqref{st:eff} and hence $\tilde\epsilon$ cannot be associated to its eigenvalue as for usual neutral fluids. However, it can be defined by \emph{subtracting} the boost vector such that $(\tilde T^{ab}+\Xi\tilde{\mathcal{P}}_\Xi\tilde v^{a}\tilde v^{b})\tilde u_{b}=-\tilde\epsilon \tilde u^{a}$. Given these definitions, the boosted fluid satisfies the expected local thermodynamic relations consistent with the first law,
\beq \label{gibbs2}
\begin{split}
d\tilde P=\tilde sd\tilde{\mathcal{T}}+\sum_{a=1}^{k}\tilde{\mathcal{J}}_{(a)}d\tilde \omega_{a}+\tilde{\mathcal{P}}_{\Xi}d\Xi~~,~~d\tilde \epsilon=\tilde{\mathcal{T}}d\tilde s+\sum_{a=1}^{k}\tilde \omega_{a}d\tilde{\mathcal{J}}_{(a)}+\Xi d\tilde{\mathcal{P}}_{\Xi}~~.
\end{split}
\eeq

We now focus on stationary configurations and solve the conservation equation \eqref{eq:bfeom1} using \eqref{gibbs1}-\eqref{gibbs2}. It is easy to see that if the boost velocities $\tilde u^{a}$ are aligned with a worldvolume Killing vector field $\tilde\bk^{a}$ and the local temperature and angular velocities satisfy $\tilde{\mathcal{T}}=\tilde T/\tilde \bk$ and $\tilde{\omega}_a=\tilde \Omega_a/\tilde \bk$ then the conservation equations reduce to
\beq \label{cons1}
\tilde{\mathcal{P}}_{\Xi}\nabla^{b}\Xi+\Xi\tilde{\mathcal{P}}_{\Xi}\tilde{\mathfrak{a}}^{b}-\nabla_{a}\left(\Xi\tilde{\mathcal{P}}_{\Xi}\tilde v^{a}\tilde v^{b}\right)=0~~.
\eeq
Moreover, we take $\tilde v^{a}$ to be proportional to the spatial worldvolume Killing vector field and we parametrize it such that $\tilde v^{a}=\tilde{\textbf{v}}^{a}/\tilde{\textbf{v}}$. Then the choice $\Xi=\tilde{\textbf{v}}/\tilde\bk$ solves \eqref{cons1} together with the form of the vector $\tilde{\textbf{v}}^{a}$,
\beq \label{vector}
\tilde{\textbf{v}}^a\partial_a=\Omega\partial_\phi~~.
\eeq
This gives rise to the effective action \eqref{eff:2}. It is worth mentioning that \eqref{vector} is valid for curved backgrounds for which $R_0\ne1$ as well as the effective action \eqref{eff:2} as long as the length scales associated with background curvatures are much larger than $\tilde\omega_a^{-1}$. As mentioned previously, this underlying theory does not have a rest frame and hence the vector $\tilde v^{a}$ must always be non-zero and take components in the helicoidal direction $\phi$.

As mentioned above, configurations built from this local effective theory with local Gibbs free energy \eqref{localgibbs} may in principle preserve $k$ $U(1)$ symmetries associated with each of the transverse angular momenta and a set of $\tilde p$ spatial translations associated with the currents $\tilde T^{ab}\tilde{\textbf{k}}_{b}$. However, the effective geometry \eqref{effectivestring} obtained from the helicoid by integrating out the finite interval $\mathbb{I}$ only preserves a $U(1)$ family of spatial isometries whose orbits are not closed. From the point of view of the effective theory, this means that only a linear combination of the translational symmetries and the set of $k$ $U(1)$ symmetries is preserved. In order to obtain the effective theory that leads to the helicoid geometry \eqref{effectivestring} we need to break the symmetries of the underlying effective theory by making the following global identification
\beq\label{req:hell}
\tilde \Omega_a =\frac{\tilde{\textbf{v}}}{\tilde R_{\phi}}~~,
\eeq
where $\tilde R_{\phi}$ is the modulus of the Killing vector field $\partial_\phi$ along the helicoidal direction $\phi$. In practice, the requirement \eqref{req:hell} implies that $\tilde \Omega_a=\Omega$. This means that a helical string can be seen as a boosted straight string with transverse angular momenta in which the transverse angular velocities and the boost velocity are related to each other. 

As we do not have a first principle derivation of the underlying effective theory after integrating out the degrees of freedom, since we do not know at the moment how to implement the procedure of Sec.~\ref{sec:int} directly in the metric \eqref{ds:blackp}, we are not able to extract with certainty all the information regarding the local effective theory of helicoidal branes. Therefore, wether or not the effective theory presented here holds when \eqref{req:hell} is not imposed is an open question, which would be worthwhile exploring.

\subsubsection*{Effective theory for helicoidal $p$-branes}
The effective free energy functional \eqref{eff:2} was derived for $\tilde p=1$. However, there are several ways in which we can generalise it for $\tilde p\ge1$. We focus first on the simplest one, which is to add $l$ flat directions to the geometry \eqref{ds:helicalp} and boost them with boost velocities $\beta_a~,~a=1,..,l$ along the flat $x_a$-directions. The resulting effective geometry is 
\beq \label{eff:try}
\tilde{\textbf{ds}}^2=-d\tau^2+\lambda^2d\phi^2+\sum_{a=1}^{l}dx_a^2~~,~~\tilde{\bk}^{a}\partial_a=\partial_\tau+\Omega\partial_\phi+\sum_{a=1}^{l}\beta_a\partial_{x_a}~~,
\eeq
and has topology $\mathbb{R}^{l+1}\times\mathbb{S}^{D-l-3}$. Therefore we refer to these configurations as \emph{helicoidal black $p$-branes}.
The resultant effective theory, which was described in the previous section for $\tilde p=1$ and holds for all $\tilde p\ge1$, is not spatially isotropic because the helicoidal direction $\phi$ is genuinely different from the remaining worldvolume directions since in that direction we cannot take the boost to be zero. The resultant effective free energy functional takes the same form of \eqref{eff:2} with the same vector \eqref{vector}.

Another generalisation of the effective theory described above can be obtained by introducing more helicoidal directions \cite{Armas:2015kra}. Consider, for example, embedding an $l$ number of the two-dimensional helicoids of Sec.~\ref{sec:helicoids} in $\mathbb{R}^{3l}$. These configurations solve the blackfold equations \eqref{eq:bfeom1} and have induced geometry and Killing vector field given by
\beq \label{geo2}
\ds^2=-d\tau^2+\sum_{a=1}^{l}\left(d\rho_{a}^2+(\lambda_a^2+a_a^2\rho_a^2)d\phi_a^2\right)~~,~~\bk^{a}\partial_a=\partial_\tau+\sum_{a=1}^{l}\Omega_a\partial_{\phi_a}~~.
\eeq
Therefore, boundaries appear when $\sum_{a=1}^l\Omega_a^2(\lambda_a^2+a_a^2\rho_a^2)=1$. The free energy functional \eqref{eq:free} takes the form
\beq \label{geoeff}
\mathcal{F}=\frac{\Omega_{(n+1)}}{16\pi G}r_+^{n}\int d\phi\int d\rho \left(\prod_{a=1}^{l}\sqrt{\lambda_a^2+a_a^2\rho_a^2}\right)\left(1-\sum_{a=1}^l\Omega_a^2(\lambda_a^2+a_a^2\rho_a^2)\right)^{\frac{n}{2}}~~,
\eeq
where $d\phi=\prod_{a=1}^{l}d\phi_a$ and $d\rho=\prod_{a=1}^{l}d\rho_a$. Adding extra flat directions to \eqref{geo2} and integrating the free energy \eqref{geoeff} gives rise to an effective action for helicoidal branes with $l$ helicoidal directions and with topology $\mathbb{R}^{(l)}\times\mathbb{S}^{(D-2-l)}$. The limit $\lambda_a\to0,\forall a$ leads to a $2l$ even-ball geometry which describes Myers-Perry black holes with several ultraspins. While such generalisation would certainly be of interest, we leave the specific details for future work.

\subsection{Thermodynamic properties and stability}  
In this section we collect the thermodynamic formulae for the conserved charges of stationary configurations constructed from helicoidal black branes and analyse the stability properties of these branes under perturbations along transverse directions. Prior to imposing \eqref{req:hell}, the effective theory is characterised by $k$ conserved particle currents of transverse angular momentum. Furthermore, as we are dealing with stationary fluid configurations the entropy current $\tilde J_{s}=\tilde s u^{a}$ must also be conserved. Integrating these currents leads to the total entropy $\tilde S$ and transverse angular momenta $\tilde{J}_{(b)\perp}$, given by
\beq
\begin{split}
\tilde S=&\frac{V_{(\tilde n+1)}}{4G}\frac{\tilde r_+^{\tilde n-k}}{\prod_{a=1}^{k}\tilde\Omega_a}\int_{\mathcal{B}_{\tilde p}}dV_{(\tilde p)}~\tilde R_0~\tilde \bk^{\tilde n} f~~,\\
\tilde{J}_{(b)\perp}=&\frac{V_{(\tilde n+1)}}{16\pi G}\frac{\tilde r_+^{\tilde n-k}}{\tilde\Omega_b\prod_{a=1}^{k}\tilde\Omega_a}\int_{\mathcal{B}_{\tilde p}}dV_{(\tilde p)}~\tilde R_0~\tilde \bk^{\tilde n}f~~, \\
\end{split}
\eeq
where we have again assumed that $\sqrt{-\gamma}d^{\tilde p+1}\tilde\sigma=\tilde R_0dV_{(\tilde p)}$ and introduced the short notation $f=f(-\frac{1}{2},k,\tilde n,\frac{\tilde\bk^2}{\tilde{\textbf{v}}^2})$. Furthermore, we also have the total mass $M$ of the system, angular momenta $\tilde J_{(b)}$ along worldvolume directions (which are not helicoidal) and the angular momentum $\tilde J_\phi$ along the helicoidal direction $\phi$ associated with the currents $\tilde T^{ab}\tilde{\textbf{k}}_{b}$. These are given by,
\beq
\begin{split}
\tilde M=&\frac{V_{(\tilde n+1)}}{16\pi G}\frac{\tilde r_+^{\tilde n-k}}{\prod_{a=1}^{k}\tilde\Omega_a}\int_{\mathcal{B}_{\tilde p}}dV_{(\tilde p)}~\tilde R_0~\tilde \bk^{\tilde n} f\left(1+\frac{\tilde R_0^2}{\tilde \bk^2}\left(\tilde n+k \frac{\tilde \bk^2}{2\tilde{\textbf{v}}^2}\frac{f'}{f}\right)\right)~~,\\
\tilde J_{(b)}=&\frac{V_{(\tilde n+1)}}{16\pi G}\frac{\tilde r_+^{\tilde n-k}\Omega_{b}}{\prod_{a=1}^{k}\tilde\Omega_a}\int_{\mathcal{B}_{\tilde p}}dV_{(\tilde p)}~\tilde R_0~\tilde R_b^2~\tilde \bk^{\tilde n-2}f\left(\tilde n +k\frac{\tilde \bk^2}{2\tilde{\textbf{v}}^2}\frac{f'}{f}\right)~~,\\
\tilde J_\phi=&\frac{V_{(\tilde n+1)}}{16\pi G}\frac{\tilde r_+^{\tilde n-k}\Omega}{\prod_{a=1}^{k}\tilde\Omega_a}\int_{\mathcal{B}_{\tilde p}}dV_{(\tilde p)}~\tilde R_0~\tilde R_\phi^2\tilde \bk^{\tilde n-2}f\left(\tilde n+k\frac{\tilde\bk^2}{2\tilde{\textbf{v}}^2}\frac{f'}{f}\right)~~,
\end{split}
\eeq
where we have defined $f'=f(\frac{1}{2},k+2,\tilde n+2,\frac{\tilde\bk^2}{\tilde{\textbf{v}}^2})$. The angular velocities $\Omega_b$ are the angular velocities along non-helicoidal worldvolume directions and $\tilde R_{b}$ the modulus of the corresponding Killing vector fields $\tilde \chi_{(b)}$ while $\Omega$ is the angular velocity along the helicoidal direction and $\tilde R_{\phi}$ is the modulus of the associated Killing vector field $\tilde \chi_\phi=\partial_\phi$. When imposing condition \eqref{req:hell}, the charges $\tilde{J}_{(b)\perp}$ and $\tilde J_\phi$ are not independently conserved but instead the linear combination $\tilde J_\phi+\sum_{b=1}^{k}\tilde{J}_{(b)\perp}$ is conserved. However, they still retain their original physical meaning, namely, they quantify the angular momentum on each of the background two-planes where the configuration is rotating.

\subsubsection*{The speed of elastic waves}
\begin{figure}[h!] 
\centering
  \includegraphics[width=0.5\linewidth]{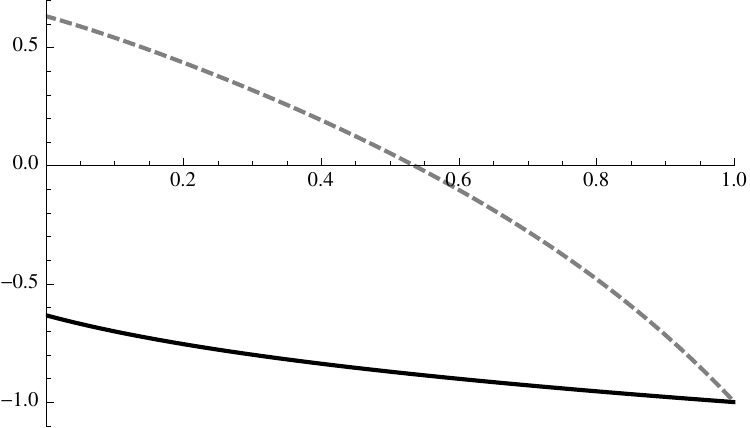}
  \begin{picture}(0,0)(0,0)
\put(-240,105){ $c_{T}^{\pm}$}  
\put(-25,65){ $ \beta $}
\end{picture} 
\caption{The speed of elastic waves for $\tilde n=2$ and $k=1$. The upper branch (dashed gray) corresponds to $c_{T}^{+}$ and the lower branch (solid black) corresponds to $c_{T}^{-}$.} \label{fig:elastic}
\end{figure}
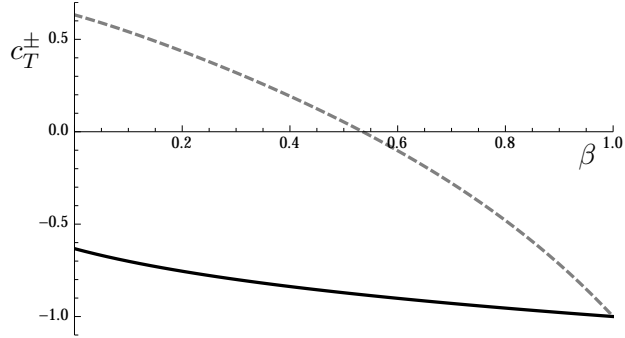

Here we study the stability properties of this class of branes restricting the analysis to transverse long-wavelength perturbations. We focus on the case of a two-dimensional worldvolume geometry, so that spatial anisotropic effects are not present. We consider a brane trivially embedded into flat space-time such that its induced metric is the two-dimensional Minkowski metric $\tilde\eta_{ab}$. Since the fluid does not have a rest frame, we consider a fluid velocity of the form $\tilde u^{a}\partial_a=(\partial_\tau+\beta\partial_\phi)/(\sqrt{1-\beta^2})$. Hence we have that $\tilde v^a\partial_a=\partial_\phi$. We then perform small perturbations in the local thermodynamic potentials and in the boost velocity $\beta$ and deform the embedding geometry by a small amount $\tilde X^{i}\to\tilde X^{i}+\tilde\varepsilon^{i}$. The extrinsic curvature in turn changes to ${{\tilde{K}_{ab}}}^{i}\to{{\tilde{K}_{ab}}}^{i}+\partial_a\partial_b\tilde\varepsilon^{i}$. Considering a long-wavelength perturbation of the form $\tilde\varepsilon^{i}=\tilde A^{i} e^{i(w\tau+\kappa \phi)}$, the extrinsic equation of motion takes the form
\beq
\left((\tilde\epsilon+\beta^2\tilde P)w^2+(\tilde P+\beta^2\tilde\epsilon-\Xi\tilde{\mathcal{P}}_{\beta}(1-\beta^2))\kappa^2+2w\kappa(\tilde\epsilon+\tilde P)\beta\right)\tilde\varepsilon^{i}=0~~.
\eeq
The resultant dispersion relation can be expressed as $w=c_{T}^{\pm}\kappa$, that is, there are two branches with different speeds of propagation of elastic waves. In general $c_{T}^{\pm}$ is always real and has no imaginary part. In Fig.~\ref{fig:elastic} we show the behaviour of these two branches as a function of $\beta$ for $\tilde n=2$ and $k=1$.
The behaviour is generic for all values of $\tilde n$ and $k$. We therefore see that these branes are elastically stable under small perturbations. In practice, this means that the helicoidal black rings which we will construct in Sec.~\ref{sec:helicoidalrings} are stable under small extrinsic perturbations.

%%%%%%%%%%%%%%%%%%%%%%%%%%%%%%%%%%%%%
%%%%%%%%%%%%%%%%%%%%%%%%%%%%%%%%%%%%%

\section{Doubly-spinning black rings from black toroids} \label{sec:blacktor}
In this section we apply the effective theory of ultraspining Myers-Perry branes of Sec.~\ref{sec:mpbranes} to the case of doubly-spinning black rings, which can be easily generalised to the case of several transverse angular momenta. We analyse higher-order corrections to the thermodynamics, by obtaining a linear combination of the coefficients $\tilde\lambda_i$ for this class of branes. In the end, we depict its phase diagram. 

\subsection{Leading order doubly-spinning black rings}
Since we have the effective action \eqref{eq:freestring1} we can readily use it and embed it into a ring geometry. Instead, we will show, for the purpose of exemplification, how the effective action arises for this particular case. For that purpose, consider a black toroid geometry, obtained by revolving the disc geometry \eqref{ds:disc} around the $x_2$-axis with a radius $R$. Explicitly, we can embedded the black toroid in flat space-time with coordinates $(t,x_i)$ by choosing $t=\tau$ and the mapping functions
\beq
\!\!\!\!X^{1}(\rho,\phi_1)=\rho\cos\phi_1,~X^{2}(\rho,\phi_1)=\rho\sin\phi_1,~X^{3}(R,\phi)=R\sin\phi,~X^{4}(R,\phi)=R\sin\phi~,
\eeq
leading to the induced $p=3$ worldvolume geometry
\beq \label{ds:blacktoroid}
\textbf{ds}^2=-d\tau^2+d\rho^2+\rho^2d\phi_1^2+R^2d\phi^2~~,
\eeq
where $\rho\ge0$ and $0\le\phi_1,\phi\le2\pi$. We set the geometry to rotate with angular velocities $\tilde \Omega$ and $\Omega$ on each of the Cartan angles such that
\beq
\textbf{k}^{a}\partial_a=\partial_\tau+\tilde \Omega\partial_{\phi_1}+\Omega\partial_{\phi}~~,~~\textbf{k}^2=1-\tilde \Omega^2\rho^2-\Omega^2R^2~~.
\eeq
The geometry acquires a boundary at $\textbf{k}=0$ which is somewhat modified compared to the disc, namely, the boundary is at
\beq
\rho_+=\frac{\sqrt{1-\Omega^2R^2}}{\tilde \Omega}~~,
\eeq
and hence requires that $\Omega^2R^2<1$. The brane thickness $r_0$ varies according to 
\beq
r_0=r_+\sqrt{1-\tilde \Omega^2\rho^2-\Omega^2R^2}~~,
\eeq
and thus takes a maximum value at the origin $\rho=0$ and shrinks to zero at the boundary $\rho=\rho_+$. It is clear from the geometry \eqref{ds:blacktoroid} that $(\tau,\phi)=constant$ sections have the geometry of a disc. The topology of these configurations, valid in the regime $r_+\ll \rho_+,\tilde\Omega r_+\ll 1,r_0\ll R$,  is therefore $\mathbb{S}^{1}\times\mathbb{S}^{(D-3)}$ in $D\ge7$, where $R$ measures the size of the $\mathbb{S}^{1}$ and $r_0$ the size of the $\mathbb{S}^{(D-3)}$. In particular, we see that when the size of the disc $R$ shrinks to zero, there is a topology-changing transition from $\mathbb{S}^{1}\times\mathbb{S}^{(D-3)}\to\mathbb{S}^{(D-2)}$ and we recover the disc. However, at the level of the free energies, as we shall see, the limit $R\to0$ does not lead to the free energy of Myers-Perry black holes \eqref{free:mp22}. On the other hand, the limit $R\to\infty$ and $\Omega R\to0$, while introducing a new coordinate $z=R\phi$ leads to a Myers-Perry string with topology $\mathbb{R}\times \mathbb{S}^{(D-3)}$ and induced worldvolume metric \eqref{ds:sdisc}.

Indeed, the black toroid constructed here is nothing more than the outcome of taking a string of ultraspinning Myers-Perry black holes with a single spin and bending it into a ring of radius $R$. These geometries, therefore, describe doubly-spinning black rings with two angular momenta in $D\ge7$ in the ultraspinning regime and hence are captured by the effective free energy \eqref{eq:freestring1}.

\subsubsection*{Equilibrium condition and physical properties}

We now integrate out the disc geometry $(\rho,\phi_1)$ and obtain an effective two-dimensional string geometry bent over a circle of radius $R$ with a conserved transverse angular momentum current,
\beq \label{ds:eff}
\tilde \ds^2=-d\tau^2+R^2d\phi^2~~,~~\tilde \bk^a\partial_a=\partial_\tau+\Omega\partial_\phi~~,
\eeq
and effective free energy
\beq
\tilde{\mathcal{F}}[R]=\frac{\Omega_{(\tilde n+1)}}{8G}\frac{\tilde r_+^{\tilde n-2}}{\tilde \Omega^2}R\left(1-\Omega^2R^2\right)^{\frac{\tilde n}{2}}~~,
\eeq
where we have used that $n=\tilde n-2$ since we have integrated out a two-plane. Varying this free energy with respect to $R$ and solving the resulting equation of motion we find
\beq \label{cond:br}
\Omega R=\frac{1}{\sqrt{\tilde n+1}}~~.
\eeq
This condition is independent of $\tilde\Omega$, which can be freely chosen as long as it satisfies the validity requirements of the underlying effective theory, which translate into $\tilde\Omega\tilde r_+\ll1$. The equilibrium condition \eqref{cond:br} is the same as that for singly-spinning black rings in $D\ge7$ \cite{Emparan:2009vd} and can be expressed as $\Omega R=1/\sqrt{D-3}$. Therefore, we see that in the ultraspinning regime in $D\ge7$, the equilibrium condition for doubly-spinning black rings does not receive any corrections due to an extra angular momentum, in agreement with the arguments in \cite{Armas:2011uf}.  The thermodynamic properties, which can now be derived from the free energy read
\beq \label{dss1}
\tilde M=\frac{\Omega_{(\tilde n+1)}}{8G}\frac{r_+^{\tilde n-2}}{\tilde\Omega^2}R(n+2)\left(\frac{\tilde n}{\tilde n+1}\right)^{\frac{\tilde n}{2}}~~,~~\tilde S=\frac{\Omega_{(\tilde n+1)}\pi}{2G}R\frac{r_+^{\tilde n-2}}{\tilde\Omega^2}\left(\frac{\tilde n}{\tilde n+1}\right)^{\frac{\tilde n}{2}}
\eeq
\beq \label{dss2}
\tilde J_{\perp}=\frac{\Omega_{(\tilde n+1)}}{4G}\frac{r_+^{\tilde n-2}}{\tilde\Omega^3}R\left(\frac{\tilde n}{\tilde n+1}\right)^{\frac{\tilde n}{2}}~~,~~\tilde J=\frac{\Omega_{(\tilde n+1)}}{4G}\frac{r_+^{\tilde n-2}}{\tilde\Omega^2}R^2\left(\frac{\tilde n}{\tilde n+1}\right)^{\frac{\tilde n}{2}}\sqrt{\tilde n+1}~~.
\eeq
Clearly, by taking the limit $R\to0$ in these conserved quantities leads to vanishing results. While the limit $R\to0$ in \eqref{ds:blacktoroid} leads to a disc geometry, the physical properties in this limit do not correspond to those of Myers-Perry black holes in the ultraspinning limit. As we will see in Sec.~\ref{sec:helicoidalrings}, this is not the case for helicoidal rings for which such limit exists off-shell. These geometries have also a vanishing tension, as expected, and hence satisfy the Smarr relation \eqref{eq:smarr} in asymptotically flat space-time.

We have mentioned that these geometries describe the ultraspinning limit of doubly-spinning black rings with two ultraspins. Therefore, one might be tempted to take the limit $\tilde\Omega\to0$ which supposedly would describe a singly-spinning black ring. This, however, is not correct. In fact, by looking at the thermodynamic properties above we see that $\tilde J_\perp/\tilde M\propto \tilde\Omega^{-1}$ and hence $\tilde \Omega\to0$ describes a ring with large angular momentum in the direction $\phi_1$. Therefore the size of the boundary is pushed to arbitrary large values of $\rho$ and the ring is effectively described by a $p=3$ toroidal geometry.

In order to obtain the limit in which the ring is not rotating in the $\phi_1$-direction one has to take $\tilde\Omega\to\infty$ leading to $\tilde J_\perp /\tilde M\to0$. However, this violates the validity requirement of the underlying effective theory $\tilde r_+\tilde \Omega\ll1$ and $\tilde r_+\ll R$ and hence it is not possible to send $\tilde \Omega\to\infty$ while satisfying $\tilde r_+\tilde \Omega\ll1$ and simultaneously demanding finite and non-zero conserved charges. That is, the doubly-spinning black ring must be ultraspinning in the $\phi_1$-direction for the approximation to be valid. In order to overcome this one has to consider Myers-Perry branes with arbitrary angular momentum, seen as an internal spin in the effective action \cite{Armas:2014rva}. 

\subsection{Higher-order corrections} \label{sec:hight2}
We now analyse higher-order corrections to this configuration using the methods of \cite{Armas:2013hsa, Armas:2013goa, Armas:2014bia, Armas:2014rva} to order $(\tilde r_0/R)^2$. It was shown in \cite{Armas:2013hsa} that black rings embedded in flat space-time are described by a single transport coefficient to second order, which is the linear combination $\boldsymbol{\lambda}_1=\lambda_1+\lambda_2+(1/n)\lambda_3$, appearing in the effective free energy in the manner
\beq
\int_{\mathcal{B}_p}\sqrt{-\gamma}\boldsymbol{\lambda}_1K^{i}K_{i}~~.
\eeq
Since the effective description of doubly-spinning black rings is in terms of a flat two-dimensional worldvolume \eqref{ds:eff}, these are also only characterised by one single transport coefficient to second order. Taking the original toroid configuration \eqref{ds:blacktoroid}, we have that the only non-vanishing component of the extrinsic curvature tensor is ${K_{\phi\phi}}^{i}=-R$ and hence $K^{i}=-1/R$. Integrating out the two-plane we find
\beq
\int_{\mathcal{B}_p}\sqrt{-\gamma}\boldsymbol{\lambda}_1K^{i}K_{i}=-\int_{\mathcal{B}_{\tilde p}}\sqrt{-\tilde \gamma}\left(\tilde P\tilde r_0^2\frac{(\tilde n-1)(3\tilde n-2)}{(\tilde n+2)(\tilde n-2)^2}\xi(\tilde n-2)\right)\tilde K^{i}\tilde K_{i}~~,
\eeq
where $\tilde P$ is given in \eqref{p:mp} and $\tilde {K_{ab}}^{i}={K_{ab}}^{i}$ since the two-plane did not contribute to the extrinsic curvature tensor. Therefore we read off 
\beq
\tilde{\boldsymbol{\lambda}}_1=-\tilde P\tilde r_0^2\frac{(\tilde n-1)(3\tilde n-2)}{(\tilde n+2)(\tilde n-2)^2}\xi(\tilde n-2)~~,~~\tilde n>4~~,
\eeq
which, using the map given in \cite{Armas:2013hsa}, constitutes a prediction for the Young modulus of ultraspinning Myers-Perry branes. Note that this transport coefficient is only valid for $\tilde n>4$. This is due to the fact that originally $\boldsymbol{\lambda}_1$ was valid for $n>2$ and afterwards we performed the integration over the two-plane. 

The effective free energy \eqref{eq:free}, including the second order corrections, is therefore
\beq\label{freeds}
\tilde{\mathcal{F}}[R]=-\int_{\mathcal{B}_p}\sqrt{-\gamma} \left(P+\upsilon_1\mathfrak{a}^{c}\mathfrak{a}_{c}+\boldsymbol{\lambda}_1K^{i}K_{i}\right)=-2\pi R\left(\tilde P+\frac{\tilde{\boldsymbol{\lambda}}_1}{R^2}\right)~~,
\eeq
where $\tilde P$ is the corrected pressure obtained in \eqref{pmp1}. Varying it with respect to $R$ we find the corrected equilibrium condition
\beq
\Omega R=\frac{1}{\sqrt{\tilde n+1}}\left(1+\tilde\xi(\tilde n-2)\frac{\tilde\varepsilon^2}{1-\ell^2}\right)+\mathcal{O}\left(\tilde\varepsilon^4\right)~~,
\eeq
where we have introduced the length scale associated with the underlying effective theory $\ell=\tilde r_+\tilde\Omega$ and also $\tilde\varepsilon=\tilde r_0/R$, which is the parameter controlling the order of the extrinsic derivative expansion. We have also defined for convenience the function $\tilde\xi(\tilde n-2)$ as
\beq
\tilde\xi(\tilde n-2)=\frac{(\tilde n-1)(3\tilde n-2)}{(\tilde n+2)(\tilde n-2)^2}\xi(\tilde n-2)~~.
\eeq
From here we see that, at second order in derivatives, the intrinsic angular momentum affects the equilibrium condition of doubly-spinning black rings.

\subsection{Phase diagram}
Using the thermodynamic relations \eqref{thermo1}-\eqref{thermo2} we can extract, from the free energy \eqref{freeds}, all the corrections to the conserved charges \eqref{dss1}-\eqref{dss2}. With these conserved charges we can obtain the phase structure of ultraspinning doubly-spinning back rings in $D\ge9$ by introducing the dimensionless reduced quantities $a_{\text{H}}$, $j$ and $j_\perp$ as in \cite{Emparan:2007wm}. The analytic expressions read
\beq
\begin{split}
a_{\text{H}}=\frac{4^{\frac{1}{\tilde n+1}}}{\tilde n-2}\tilde\varepsilon^{\frac{1}{\tilde n+1}}\ell^{\frac{2}{\tilde n+1}}&\left(1-\ell^2\right)^{-\frac{\tilde n+2}{\tilde n+1}}\Big((\tilde n-2)-\tilde n\ell^2 \\
&+\frac{\tilde n \left(\tilde n^2+((\tilde n-1) \tilde n-4) \ell^2+7 \tilde n+4\right)-8}{\tilde n (\tilde n+1)^2}\tilde \xi(\tilde n-2)\frac{\tilde\varepsilon ^2}{1-\ell^2} \Big)~,\\
\end{split}
\eeq
\beq
j=2^{-\frac{\tilde n+2}{\tilde n+1}}\tilde\varepsilon^{-\frac{\tilde n}{\tilde n+1}}\ell^{\frac{2}{\tilde n+1}}(1-\ell^2)^{-\frac{1}{\tilde n+1}}\left(1+\mathcal{O}\left(\tilde\varepsilon^{4}\right)\right)~~,
\eeq
\beq
j_\perp=\frac{2^{-\frac{1}{n+1}}}{\sqrt{\tilde n}}\tilde\varepsilon^{\frac{1}{\tilde n+1}}\ell^{-\frac{(\tilde n-1)}{\tilde n+1}}(1-\ell^2)^{-\frac{\tilde n+2}{\tilde n+1}}\left(1+\frac{3\tilde n+4+\tilde n(\tilde n+1)\ell^2}{\tilde n^2(\tilde n+1)(1-\ell^2)^2}\tilde\xi(\tilde n-2)\frac{\tilde\varepsilon^{2}}{1-\ell^2}\right)~,
\eeq
where we have made use the freedom to perform field redefinitions \cite{Armas:2013hsa, Armas:2013goa} of the form $R\to R+ \alpha R\tilde\varepsilon^2$, for some constant $\alpha$, in order to remove the order $\tilde\varepsilon^2$ correction from $j$.

\begin{figure}[h!] 
\centering
\includegraphics[width=0.4\linewidth]{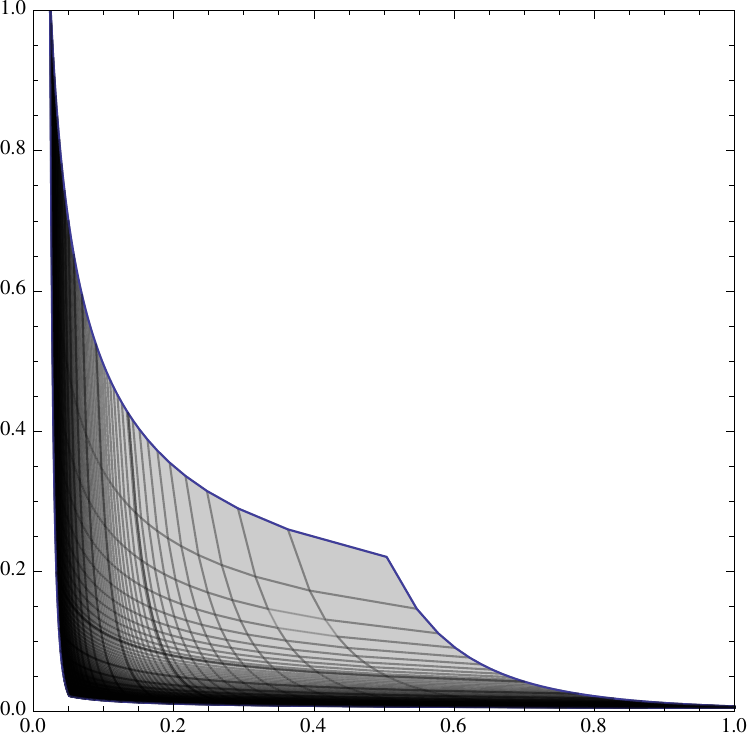}
\begin{picture}(0,0)(0,0)
\put(-190,155){ $ j $}
\put(-25,-3){ $ j_\perp $}
\end{picture} 
\caption{Phase diagram $(j,j_\perp)$ for doubly-spinning black rings in $D=9$ including second order corrections in $\ell$ and $\tilde\varepsilon$. We have rescaled both axes such that the range of reduced angular momenta lies within $0\le j,j_\perp\le1$.} \label{fig:phase}
\end{figure}
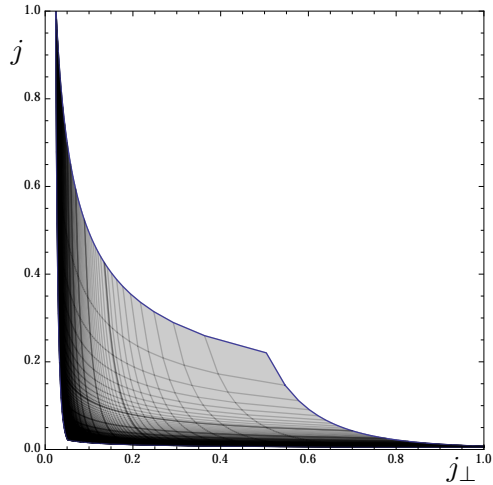

In Fig.~\ref{fig:phase} we plot the phase diagram $(j,j_\perp)$ for the range $10^{-3}\le \tilde\varepsilon, \ell\le10^{-1}$ in $D=9$.
Since we have integrated out a two-plane, the second order transport coefficients are only valid for $D\ge9$. The higher-order corrections in $\tilde\varepsilon^{2}$ do not affect significantly the phase diagram $(j,j_\perp)$, in stark contrast to the diagram $a_{\text{H}}(j,j_\perp)$, where $a_{\text{H}}$ is the reduced area, as in the case of the singly-spinning black ring \cite{Armas:2014bia}. This is due to the fact that $j_\perp$ is not very sensitive to elastic corrections, instead, corrections in the effective theory, controlled by the parameter $\ell$ contribute in a significant way since these are associated to the transverse angular momentum $j_\perp$.

%%%%%%%%%%%%%%%%%%%%%%%%%%%%%%%%%%%%%
%%%%%%%%%%%%%%%%%%%%%%%%%%%%%%%%%%%%%
%%%%%%%%%%%%%%%%%%%%%%%%%%%%%%%%%%%%% 
\section{Helicoidal black holes}\label{sec:helicoidalrings}
In this section we use the effective theory of Sec.~\ref{sec:effhelicoids} in order to construct several new classes of black hole geometries which have not previously been considered in the literature. We begin with the simplest case of helicoidal black rings and then move on to the case of helicoidal black tori. 

\subsection{Helicoidal black rings}
In this section we consider helicoidal black rings in $D\ge6$ with topology $\mathbb{S}^{1}\times \mathbb{S}^{(D-3)}$. We begin by analysing the case $k=1$, that is, helicoidal black rings, which are locally black helicoids, and hence for which the effective theory, prior to imposing \eqref{req:hell}, is described by only one single transverse angular momentum. At the end of this section we study the cases $k>1$.

Consider writing a four-dimensional subspace of the background flat space-time as the product of two 2-planes, 
\beq
d\mathbb{E}^{2}_{(4)}=dr_1^2+r_1^2d\psi_1^2+dr_2^2+r_2^2d\psi_2^2~~,
\eeq
and embed the the helicoidal ring by choosing $t=\tau$, $r_1=R$, $r_2=0$ and $\psi_1=\phi$. The effective string geometry bent over a ring of radius $R$ with Killing vector field $\tilde \bk^a$ and boost vector $\tilde{\textbf{v}}^{a}$ along the helicoidal direction $\phi$ is thus,
\beq \label{eff:helhel}
\tilde \ds^2=-d\tau^2+R^2d\phi^2~~,~~\tilde \bk^{a}\partial_a=\partial_\tau+\Omega\partial_\phi~~,~~\tilde{\textbf{v}}^{a}\partial_a=\Omega\partial_\phi~~.
\eeq
Imposing \eqref{req:hell} and since we have $k=1$, this corresponds to a black hole horizon geometry with background Killing vector field
\beq
k^{\mu}\partial_\mu=\partial_t+\Omega\partial_{\psi_1}+\Omega\partial_{\psi_2}~~,
\eeq
The direction $\phi$ is the helicoidal direction which is associated with the background two-plane with angular coordinate $\psi_1$. The rotation on the transverse two-plane (with respect to the effective geometry \eqref{eff:helhel}) associated with the angular coordinate $\psi_2$ is related to the local angular velocity $\tilde \omega_a$ of the underlying effective theory. Therefore, helicoidal rings preserve only a linear combination of the two rotational Killing vector fields and in $D=6$ they saturate the rigidity theory (with the present embedding).  It is clear that in the regime $\tilde r_0\ll R$ the geometry \eqref{eff:helhel} is locally a black helicoid, since by sending $R\to\infty$ and introducing a non-compact coordinate $\lambda \tilde \phi=R\phi$ one obtains the effective geometry of a helicoid \eqref{effectivestring}.

\begin{figure}[h!]
\centering 
\begin{subfigure}{.5\textwidth}
  \centering 
  \includegraphics[width=.8\linewidth]{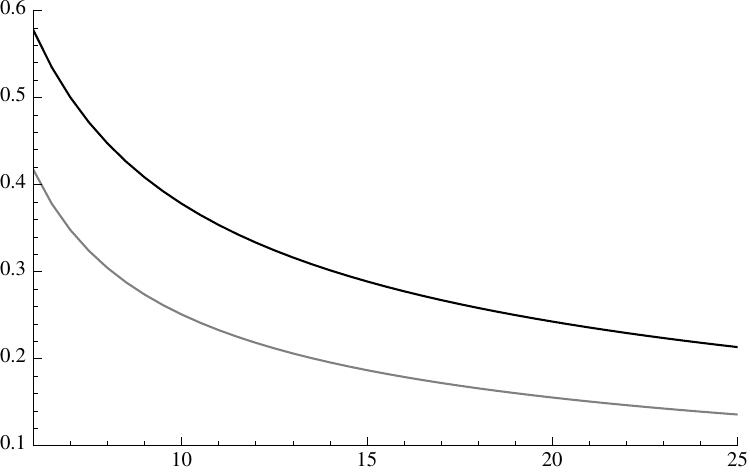}
  \begin{picture}(0,0)(0,0) 
\put(-205,90){ $\Omega R$}
\put(-28,-5){ $ D $}
\end{picture}	
\end{subfigure}% 
\begin{subfigure}{.5\textwidth}
  \centering
  \includegraphics[width=.8\linewidth]{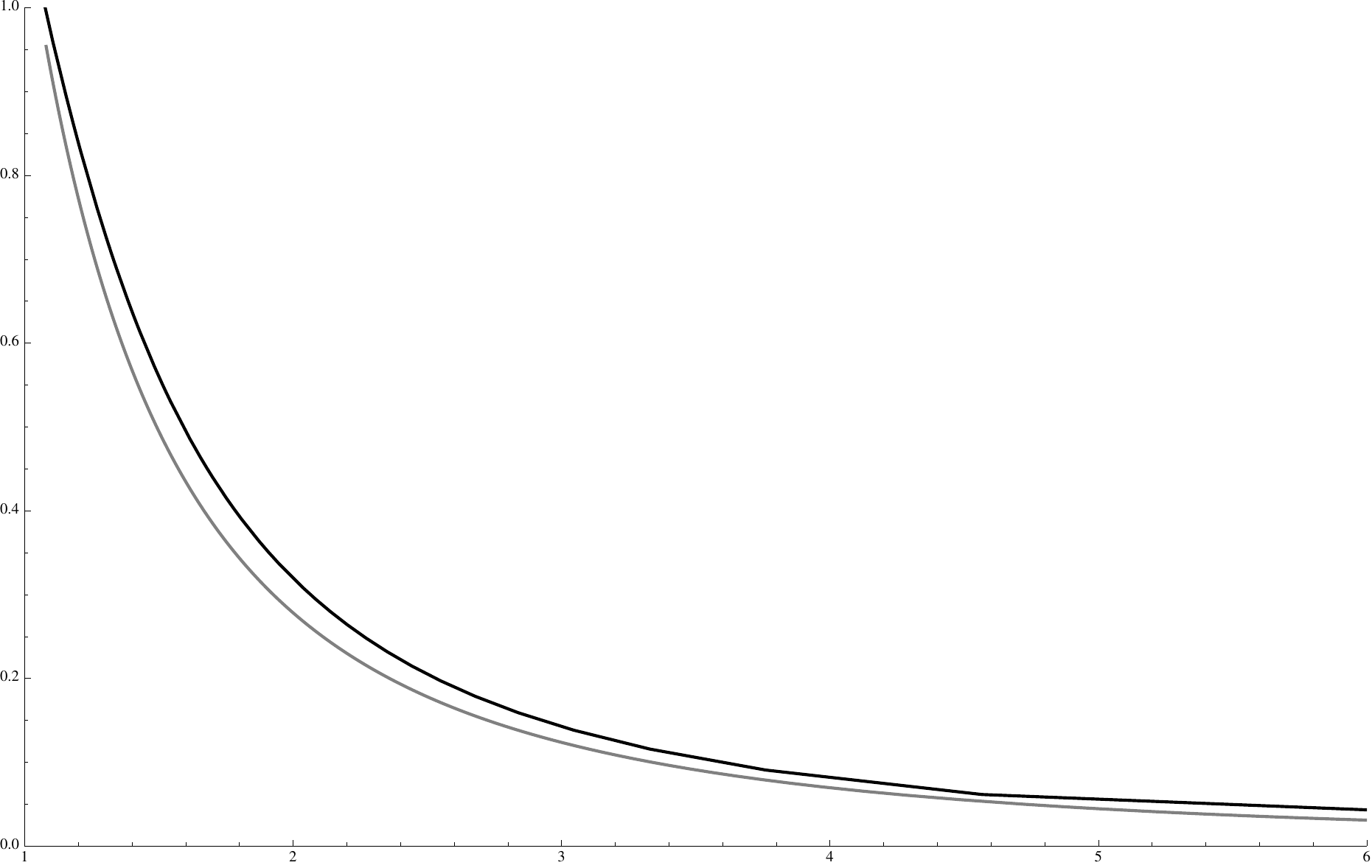}  
  \begin{picture}(0,0)(0,0)
  \put(-205,90){ $ a_{\text{H}}  $}
  \put(-28, -5){ $j$}
  \end{picture}	
\end{subfigure}
\caption{On the left we give the balance condition $\Omega R$ for helicoidal rings (gray line) as a function of the space-time dimension $D$ for $D\ge6$ while the black line represents the balance condition for black rings. On the right we have plotted the reduced area $a_{\text{H}}$ as a function of the reduced total angular momentum $j$ for helicoidal rings in $D=6$ (gray line) as well as the corresponding curve for singly-spinning Myers-Perry black holes (black line).} \label{fig:helicoidalplot}
\end{figure}
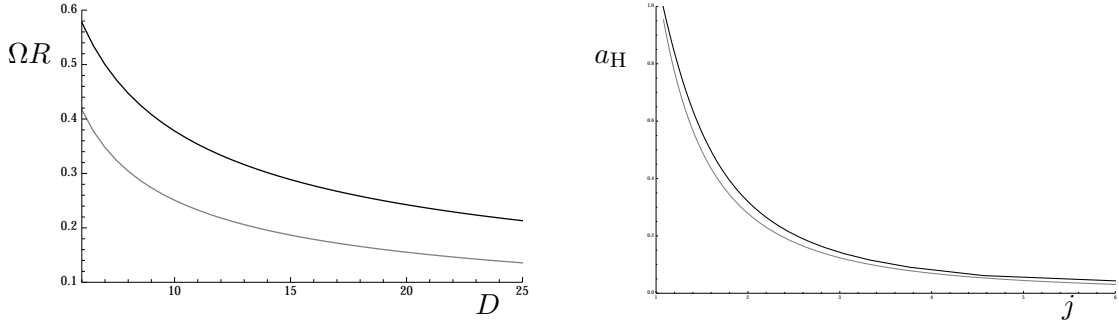

Using \eqref{eff:2}, we obtain the free energy 
\beq \label{free:helicoidalring}
\tilde{\mathcal{F}}[R]=\frac{V_{(\tilde n+1)}}{8 G}\frac{\tilde r_+^{\tilde n-1}}{\Omega}~R~\left(1-\Omega^2R^2\right)^{\frac{\tilde n}{2}}\thinspace _2\tilde{F}_1\left(-\frac{1}{2},\!\frac{1}{2};\!\frac{\tilde n+2}{2};-\frac{1-\Omega^2R^2}{\Omega^2R^2}\right)~~.
\eeq
We can now vary this free energy with respect to $R$ in order to obtain the equilibrium condition
\beq \label{eq:balance}
\frac{1-(\tilde n+1) \Omega ^2R^2}{ \left(1- \Omega ^2R^2\right)}-\frac{\,
   _2\tilde{F}_1\left(\frac{1}{2},\frac{3}{2};\frac{\tilde n+4}{2};-\frac{1-\Omega^2R^2}{\Omega^2R^2}\right)}{2 \Omega ^2R^2 \,
   _2\tilde{F}_1\left(-\frac{1}{2},\frac{1}{2};\frac{\tilde n+2}{2};-\frac{1-\Omega^2R^2}{\Omega^2R^2}\right)}=0~~.
\eeq
Comparison of this condition with the tension \eqref{eq:tensionhelicoid} leads to the conclusion that balanced helicoidal rings have a vanishing tension, as they constitute examples of compact asymptotically flat black holes. It is worth mentioning that condition \eqref{eq:balance} differs from that obtained for the usual black rings due to the existence of the second term involving the hypergeometric functions. The transcendental Eq.~\eqref{eq:balance} does not admit a solution in closed form so it has to be solved numerically. Since Eq.~\eqref{eq:balance} only depends on the combination $\Omega R$ we present its solution (gray line) in Fig.~\ref{fig:helicoidalplot} as a function of the space-time dimension $D$, valid in the regime $\Omega\tilde r_+\ll1$ and $\tilde r_0\ll R$, along with the balance condition $\Omega R=1/\sqrt{D-3}$ for black rings (black line) \cite{Emparan:2007wm}. We see that for a given radius $R$ and dimension $D$, helicoidal rings need to rotate slower in order to reach equilibrium.

\subsubsection*{The Myers-Perry limit}
The thermodynamic properties of these configurations can be obtained using Eqs.~\eqref{thermo1}-\eqref{thermo2}. Alternatively, they can be obtained using \eqref{eq:masshell}-\eqref{eq:anghell2} by setting $\lambda=R$ and integrating over $\phi$ in the range $0\le\phi\le2\pi/a$ and setting $a=1$. As in the case of the black helicoids, the free energy \eqref{free:helicoidalring} reduces to the singly-spinning Myers-Perry black hole when taking $R\to0$. This however, for the case of helicoidal black rings, is only true off-shell, that is, without imposing the balancing condition \eqref{eq:balance}. In order to see that this is indeed the case, we have evaluated the reduced quantities $a_{\text{H}}$ and $j$, where $j$ is the total angular momentum of the helicoidal black ring, and compared it in Fig.~\ref{fig:helicoidalplot} to singly-spinning Myers-Perry black holes in $D=6$. The blackfold approximation is expected to hold, in this case, in the regime $\tilde r_+ \Omega \ll1$ and $\tilde r_0\ll R$, which is equivalent to $j\gg1$. In Fig.~\ref{fig:helicoidalplot} we have extrapolated the curve $a_{\text{H}}(j)$ to values of $j\sim\mathcal{O}(1)$. We note that we have rescaled the free energy such that $\mathcal{\tilde F}\to(1/2)\mathcal{\tilde F}$, as explained in Sec.~\ref{sec:helicoids}, in order to compare with the limit $R\to0$. The actual curve $a_{\text{H}}(j)$, without the rescaling, is placed in a slightly lower position compared to the phase diagram of Myers-Perry black holes in Fig.~\ref{fig:helicoidalplot}.

As it can be seen from Fig.~\ref{fig:helicoidalplot}, the curve $a_{\text{H}}(j)$ for helicoidal rings does not cross the corresponding curve for singly-spinning Myers-Perry black holes. However, if one holds the angular velocity $\Omega$ fixed and takes the limit $R\to0$, this results in the Myers-Perry geometry in the ultraspinning regime. This indicates that it would be necessary to move along a curve of unbalanced helicoidal rings. It would be interesting to understand the role of these solutions in the phase structure of higher-dimensional black holes with one single angular momentum. 

\subsubsection*{Higher-order corrections}
As in the case of doubly-spinning black rings, one may consider refining the approximation by taking into account higher-order corrections. However, in this case we do not have enough information regarding the necessary transport coefficients. In particular, we find the non-vanishing contribution to the effective free energy
\beq
\int_{\mathcal{B}_p}\sqrt{-\gamma}\upsilon_{2}\mathcal{R}\propto \int_{\mathcal{B}_{\tilde p}}\sqrt{-\gamma}\alpha \tilde K^{i}\tilde K_{i}~~,
\eeq
for some $\alpha$, where we have only used the fact that $\upsilon_{2}\propto r_0^{n+2}$ by dimensional analysis. Since this gives a non-vanishing contribution and we do not have information about $\upsilon_{2}$, we cannot study higher-order corrections in this case.

\subsubsection*{No helicoidal black rings with $k>1$}
Helical black rings (locally helical black strings) were constructed in \cite{Emparan:2009vd} and we have shown above that helicoidal rings (locally black helicoids) are also possible geometries for black hole horizons. We now test if helicoidal rings, locally black helicoid $k$-branes, can be found. In this case the effective theory is that of a helicoidal string with a $k$ number of particle currents. Using \eqref{eff:2} for arbitrary $k$ and varying it with respect to $R$ we obtain a simple modification of the equilibrium condition \eqref{eq:balance}, namely,
\beq \label{eq:balance2}
\frac{1-(\tilde n+1) \Omega ^2R^2}{ \left(1- \Omega ^2R^2\right)}-\frac{k\,
   _2\tilde{F}_1\left(\frac{1}{2},\frac{k+2}{2};\frac{\tilde n+4}{2};-\frac{1-\Omega^2R^2}{\Omega^2R^2}\right)}{2 \Omega ^2R^2 \,
   _2\tilde{F}_1\left(-\frac{1}{2},\frac{k}{2};\frac{\tilde n+2}{2};-\frac{1-\Omega^2R^2}{\Omega^2R^2}\right)}=0~~.
\eeq
We have explicitly checked that this condition has no solution for the range of parameters $2\le \tilde n-k\le100$ and $2\le k\le 100$. This means that for $k\ge2$ the centrifugal repulsion, induced by rotation, cannot compensate the gravitational tension of the $k$-helicoid. 

%%%%%%%%%%%%%%%%%%%%%%%%%%%%%%%%%%%%%%
%%%%%%%%%%%%%%%%%%%%%%%%%%%%%%%%%%%%%%
%%%%%%%%%%%%%%%%%%%%%%%%%%%%%%%%%%%%%%

\subsection{Helicoidal black tori}
In this section we construct helicoidal black tori. This is an anisotropic solution, in which one direction of the torus is helicoidal and the other is not. Black tori, using the blackfold effective theory \eqref{tbf}-\eqref{sbf}, have been constructed in \cite{Emparan:2009vd} and exist for all values of their codimension. The case of helicoidal black tori is rather different and in fact we were only able to find a specific codimension, namely $\tilde n=2$, which solves the blackfold equations. 

We write a six-dimensional subspace of the background flat space-time as the product of three 2-planes, 
\beq
d\mathbb{E}^{2}_{(6)}=dr_1^2+r_1^2d\psi_1^2+dr_2^2+r_2^2d\psi_2^2+dr_3^2+r_3^2d\psi_3^2~~,
\eeq
and embed the black tori by choosing $t=\tau$, $r_1=R_\phi$, $r_2=R_2$, $r_3=0$ and $(\psi_1,\psi_2)=(\phi,\phi_2)$ such that the resulting worldvolume geometry is given by
\beq
\tilde \ds^2=-d\tau^2+R_\phi^2d\phi^2+R_2^2d\phi_2^2~~,
\eeq
where $0\le\phi,\phi_2\le2\pi$. This is the standard embedding of the Clifford torus. We set it to rotate with angular velocities $\Omega$ and $\Omega_2$ and choose the direction $\phi$ to be the helicoidal direction. Therefore,
\beq
\tilde \bk^{a}\partial_a=\partial_\tau+\Omega\partial_{\phi} +\Omega_2\partial_{\phi_2} ~~,~~\tilde{\textbf{v}}^{a}\partial_a=\Omega\partial_{\phi}~~.
\eeq
This corresponds to a background Killing vector field of the form
\beq
k^{\mu}\partial_\mu=\partial_t+\Omega\partial_{\psi_1}+\Omega_{2}\partial_{\psi_2}+\Omega\partial_{\psi_3}~~,
\eeq
where we have imposed \eqref{req:hell}. Here, $\psi_3$ is the angular coordinate on the transverse two-plane associated with $\tilde\omega_a$ in the effective theory. The free energy \eqref{eff:2} becomes
\beq
\tilde{\mathcal{F}}[R_\phi,R_2]=\frac{V_{(\tilde n+1)}}{16\pi G}\frac{\tilde r_+^{\tilde n-k}}{\Omega^{k}}4\pi^2~R_\phi R_2~\tilde{\bk}^{\tilde n}\thinspace _2\tilde{F}_1\left(-\frac{1}{2},\!\frac{k}{2};\!\frac{\tilde n+2}{2};-\frac{\tilde\bk^2}{\tilde{\textbf{v}}^2}\right)~~.
\eeq
Upon variation with respect to $R_\phi$ and $R_2$ we obtain a set of two coupled equations,
\beq \label{eq:balance4}
1-\frac{n\tilde{\textbf{v}}^2}{\tilde\bk^2}-\frac{k(1-\tilde{\textbf{v}}^2)}{2\tilde{\textbf{v}}^2}\frac{\thinspace _2\tilde{F}_1\left(\frac{1}{2},\!\frac{k+2}{2};\!\frac{\tilde n+4}{2};-\frac{\tilde\bk^2}{\tilde{\textbf{v}}^2}\right)}{\thinspace _2\tilde{F}_1\left(-\frac{1}{2},\!\frac{k}{2};\!\frac{\tilde n+2}{2};-\frac{\tilde\bk^2}{\tilde{\textbf{v}}^2}\right)}=0~~,
\eeq
\beq \label{eq:balance5}
1-\frac{n\Omega_2^2R_2^2}{\tilde\bk^2}-\frac{k\Omega_2^2R_2^2}{2\tilde{\textbf{v}}^2}\frac{\thinspace _2\tilde{F}_1\left(\frac{1}{2},\!\frac{k+2}{2};\!\frac{\tilde n+4}{2};-\frac{\tilde\bk^2}{\tilde{\textbf{v}}^2}\right)}{\thinspace _2\tilde{F}_1\left(-\frac{1}{2},\!\frac{k}{2};\!\frac{\tilde n+2}{2};-\frac{\tilde\bk^2}{\tilde{\textbf{v}}^2}\right)}=0~~.
\eeq
We have not  been able to find solutions with $\tilde n>2$ and $k>1$. We have only found a numerical solution for the case $\tilde n=2$ and $k=1$ for which we need
\beq
R_\phi\Omega\sim\frac{9}{25}~~,~~R_2\Omega_2=\frac{1}{2}~~.
\eeq
Therefore these helicoidal black tori are possible black hole solutions with horizon topology $\mathbb{S}^{1}\times \mathbb{S}^{1}\times\mathbb{S}^{(D-4)}$ in $D=7$. The fact that these geometries are difficult to balance is most likely due to the anisotropy of the configuration. Developing an effective theory where all directions are helicoidal, by considering the geometry \eqref{eff:try}, would most likely allow us to find helicoidal tori in any dimension.

%%%%%%%%%%%%%%%%%%%%%%%%%%%%%%%%%%%%%%
%%%%%%%%%%%%%%%%%%%%%%%%%%%%%%%%%%%%%%
%%%%%%%%%%%%%%%%%%%%%%%%%%%%%%%%%%%%%%
\section{Discussion}\label{sec:conclusions}
In this paper we have presented two different classes of worldvolume effective theories for black branes and their respective effective actions when restricting to stationary configurations. Our work generalises the blackfold approach for higher-dimensional black holes \cite{Emparan:2009cs, Emparan:2009at} in which the dynamics of the fluid is integrated out in certain spatial sections of the worldvolume geometry. 

One of the worldvolume effective theories that we have obtained by integrating out disc sections turned out to describe the dynamics of the effective fluid living on a Myers-Perry brane in the ultraspinning regime. However, the second effective theory that we studied, obtained by integrating out finite line segments, is that of an effective fluid living on a helicoidal black brane - a solution of Einstein vacuum equations which is not known analytically and its existence was recently predicted in \cite{Armas:2015kra}.

We have used these effective theories to study and construct new black hole geometries such as doubly-spinning black rings in $D\ge7$ and helicoidal black rings in $D\ge6$ as well as helicoidal black tori in $D\ge7$ in asymptotically flat space-time. Considering other background space-times such as (Anti)-de Sitter and plane-waves would certainly be of interest. A preliminary study lead us to the conclusion that helicoidal black rings exist in (Anti)-de Sitter and plane-wave backgrounds and can in certain cases be static. 

It is likely that more effective theories can be built by integrating other sections of the worldvolume geometry which we did not consider here. We have noticed that if these sections are Euclidean minimal surfaces such as planes and helicoids, the dynamics of the fluid decouple and suitable compact spaces can be integrated out. Other interesting effective theories could potentially be obtained if more minimal surfaces in higher-dimensional Euclidean space were known.

In general, the method developed here provides a map between, at least, three different linearised solutions of Einstein equations: the black brane \eqref{ds:blackp}, the Myers-Perry brane and the helicoidal black brane. This map we exploited here and showed that it could be used to find relations between transport coefficients in the different effective theories. In this way, we were able to obtain the hydrodynamic transport coefficient $\upsilon_1$ for the black branes \eqref{ds:blackp} via a simple integration, which would otherwise require perturbing the black brane \eqref{ds:blackp} to second order in derivatives and solve Einstein equations. We were also able to make predictions regarding the coefficient $\tilde \upsilon_1$ for Myers-Perry branes as well as a contribution to the Young modulus for this class of branes. 

It would be interesting to understand how this map between solutions works directly at the level of the metric \eqref{ds:blackp} and how the integration can/should be performed. One then may wonder if these ideas can be applied in a time-dependent setting, hence allowing for a map between dissipative transport coefficients. Finally, one may explore whether or not this map works at the non-linear level and hence if it provides a solution generating technique. These directions are currently under investigation. 

 %%%%%%%%%%%%%%%%%%%%%%%%%%%%%%%%%%%%%%
  %%%%%%%%%%%%%%%%%%%%%%%%%%%%%%%%%%%%%%
   %%%%%%%%%%%%%%%%%%%%%%%%%%%%%%%%%%%%%%
  %%%%%%%%%%%%%%%%%%%%%%%%%%%%%%%%%%%%%%

\section*{Acknowledgements}
The work in this paper started out as an exploration of the comment (footnote 5) by Roberto Emparan. We are therefore grateful for his comments on a earlier draft of related work \cite{Armas:2015kra}. We would also like to thank I. Amado, J. Bhattacharya, J. Camps, O. J. C. Dias, J. Gath and S.S. Gopalakrishnan for useful discussions. This work has been supported by the Swiss National Science Foundation and the �Innovations- und Kooperationsprojekt C-13� of the Schweizerische Universit\"{a}tskonferenz SUK/CUS. JA acknowledges the current support of the ERC Starting Grant 335146 \textbf{HoloBHC}. JA would like to thank NBI for hospitality during several stages of this project. 
 
   %%%%%%%%%%%%%%%%%%%%%%%%%%%%%%%%%%%%%%
  %%%%%%%%%%%%%%%%%%%%%%%%%%%%%%%%%%%%%%

 %%%%%%%%%%%%%%%%%%%%%%%%%%%%%%%%%%%%%%
%%%%%%%%%%%%%%%%%%%%%%%%%%%%%%%%%%%%%%%%%%%%%%%%%%%%%%

  %%%%%%%%%%%%%%%%%%%%%%%%%%%%%%%%%%%%%%%
  %%%%%%%%%%%%%%%%%%%%%%%%%%%%%%%%%%%%%%

  %%%%%%%%%%%%%%%%%%%%%%%%%%%%%%%%%%%%%%
  %%%%%%%%%%%%%%%%%%%%%%%%%%%%%%%%%%%%%%

%%%%%%%%%%%%%%%%%%%%%%%%%%%%%%%%%%%%%%

  %%%%%%%%%%%%%%%%%%%%%%%%%%%%%%%%%%%%%%%
  %%%%%%%%%%%%%%%%%%%%%%%%%%%%%%%%%%%%%%

\addcontentsline{toc}{section}{References}
\footnotesize
\providecommand{\href}[2]{#2}\begingroup\raggedright\endgroup

\end{document}